\documentclass[twocolumn]{aastex631}

\usepackage{float}
\usepackage{threeparttable}
\usepackage{enumerate}
\usepackage{enumitem}
\usepackage{CJK}
\usepackage{ulem}

\received{December 27, 2022}
\revised{June 8, 2023}
\accepted{June 8, 2023}

\submitjournal{AJ}

\shorttitle{Special nova-like J2043+3413}
\shortauthors{X.Li et al.}

\graphicspath{{./}}

\begin{document}

\begin{CJK*}{UTF8}{gbsn}
\title{LAMOST J2043+3413 - a Fast Disk Precession SW Sextans Candidate in Period Gap}

\correspondingauthor{Xin Li}
\email{lixin@bjp.org.cn}

\author[0000-0001-5879-8762]{Xin Li (李昕)}
\affiliation{Beijing Planetarium, Beijing Academy of Science and Technology, Beijing 100044, People's Republic of China}

\author{Xiaofeng Wang (王晓锋)}
\affiliation{Physics Department, Tsinghua University, Beijing 100084, People's Republic of China}
\affiliation{Beijing Planetarium, Beijing Academy of Science and Technology, Beijing 100044, People's Republic of China}

\author{Jiren Liu (刘纪认)}
\affiliation{Beijing Planetarium, Beijing Academy of Science and Technology, Beijing 100044, People's Republic of China}

\author[0000-0002-8321-1676]{Jincheng Guo (郭金承)}
\affiliation{Beijing Planetarium, Beijing Academy of Science and Technology, Beijing 100044, People's Republic of China}

\author{Ziping Zhang (张子平)}
\affiliation{Beijing Planetarium, Beijing Academy of Science and Technology, Beijing 100044, People's Republic of China}

\author[0000-0002-3935-2666]{Yongkang Sun (孙永康)}
\affiliation{National Astronomical Observatories, Chinese Academy of Sciences, Beiing 100101, People's Republic of China}
\affiliation{School of Astronomy and Space Science, University of Chinese Academy of Sciences, Beijing 100049, People's Republic of China}

\author{Xuan Song}
\affiliation{Beijing Planetarium, Beijing Academy of Science and Technology, Beijing 100044, People's Republic of China}

\author{Cheng Liu (刘成)}
\affiliation{Beijing Planetarium, Beijing Academy of Science and Technology, Beijing 100044, People's Republic of China}

\begin{abstract}

We present follow-up photometric observations and time-series analysis of a nova-like, SW Sextans-type, cataclysmic variable (CV) candidate, LAMOST J204305.95+341340.6 (here after J2043+3413), with Gaia G-band magnitude of 15.30 and a distance of 990 pc, which was identified from the LAMOST spectrum. The photometric data were collected with the Tsinghua-NAOC 0.8-m telescope (TNT), TESS, ZTF, and ASAS-SN. The TESS light curve reveals the presence of two prominent periods of 2.587(8) hours and 1.09(5) days, corresponding to the orbital and superorbital (precession) period, respectively. The TNT data obtained in 2020 shows a possible quasi-periodic oscillation of 1426 seconds. The precession period is about three times shorter than that of CVs with similar orbital periods, indicating an unusually fast precessing accretion disk. The ZTF data is found to show a sudden decline of $\sim0.4$ mag on MJD 58979. From the intermittent behavior of the eclipse, we deduce that J2043+3413 is an intermediate inclination system of CV, similar to V795 Her, which is also situated in the period gap.

\end{abstract}

\keywords{Cataclysmic variable stars (203) -- Nova-like variable stars (1126) -- Stellar accretion disks (1579)  -- Eclipsing binary stars (444)}

\section{Introduction} \label{sec:intro}

A cataclysmic variable (CV) consists of a white dwarf (WD) primary and a low-mass Roche-lobe filling secondary \citep[][hereafter WA95]{warner_1995}. In CVs with sufficiently weak WD magnetic fields ($B<10^{\rm 6}$G, which include classical novae, nova remnants, dwarf novae, and nova-like variables (NLs)), the material transferred from the secondary spreads into an accretion disk surrounding the primary. If the magnetic field of a CV WD is larger ($10^{\rm 6}<B\leqslant10^{\rm 7}$G), the accretion disk will be partially disrupted, and material will flow through the magnetic lines. These are called as intermediate polar (IP). For the strongest magnetic CV ($B>10^{\rm 7}$G, polar), an accretion disk can not form \citep{1990SSRv...54..195C, 2002RSPTA.360.1951C, Dai_2016}.

CVs can be discovered either during their outbursts or through their spectroscopic features. Over the last decade, the Large Sky Area Multi-Object Fiber Spectroscopic Telescope (LAMOST: \citealt{1996ApOpt..35.5155W, 2012RAA....12.1197C}) has obtained millions of stellar spectra. \citet{2020AJ....159...43H} identified 245 CV candidates, 58 of which are new discoveries, using machine learning methods. These spectroscopically identified CVs lack relevant photometric observations. In this work, we chose one of them, LAMOST J204305.95+341340.6 (J2043+3413 for brevity), for follow up photometric study.

\begin{figure}[H]
   \centering
   \includegraphics[width=0.47\textwidth,angle=0]{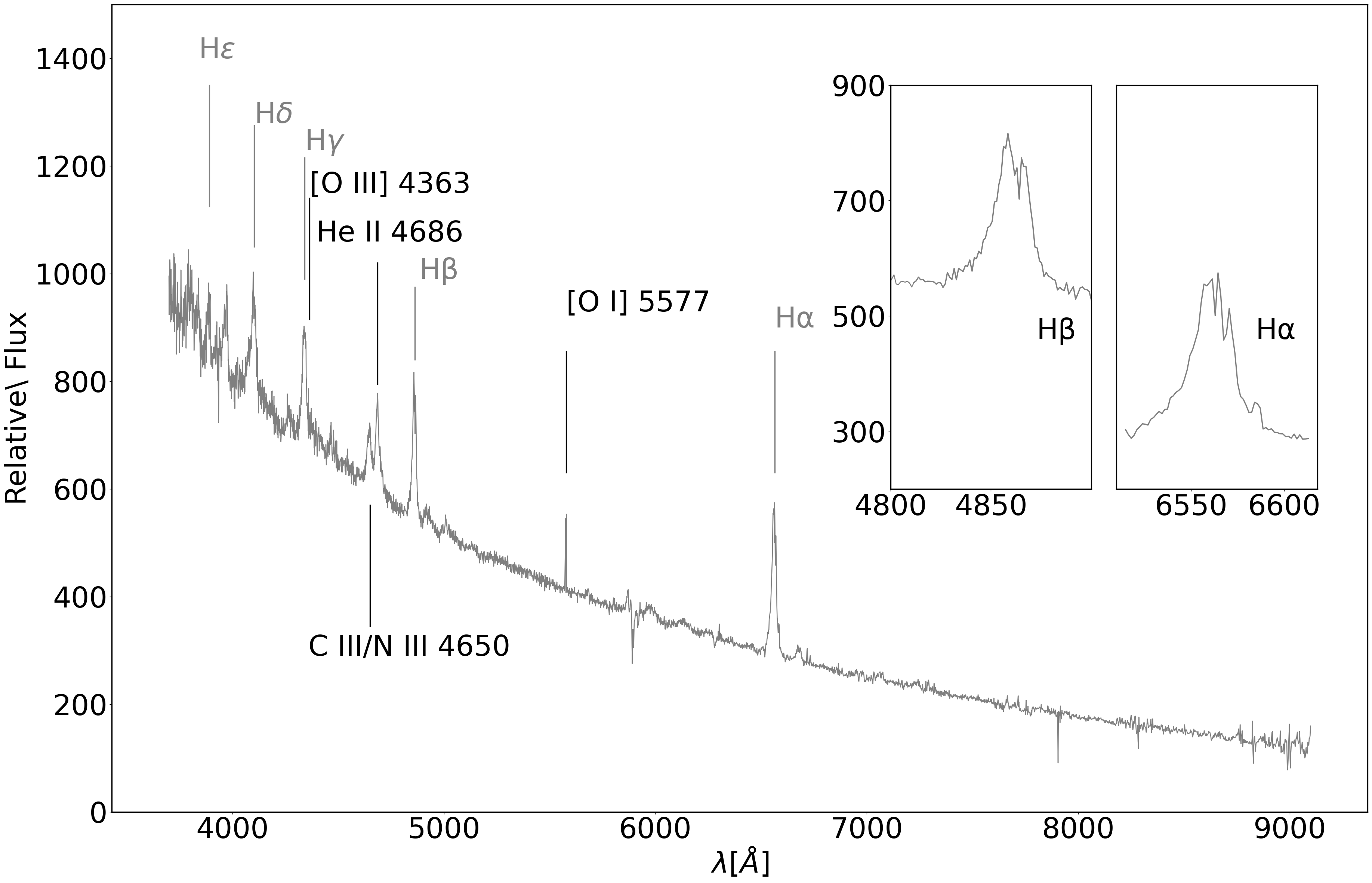}
   \caption{The LAMOST spectrum of J2043+3413 taken on MJD 57664. Balmer emission lines are marked in gray, and the others lines are marked in black. The insets show the line profiles of H$\alpha$ and H$\beta$.}
   \label{fig:J2043spectrum}
   \end{figure}

J2043+3413 was initially discovered by \citet{1984A&AS...56...87M} in a low-Galactic-latitude field survey. It has time-series archived photometry data from the Transiting Exoplanet Survey Satellite (TESS; \citealt{10.1117/1.JATIS.1.1.014003}) mission, the Zwicky Transient Facility (ZTF; \citealt{Bellm_2019, Szkody_2020, Szkody_2021}), and the All Sky Automated Survey for SuperNovae (ASAS-SN; \citealt{Kochanek_2017}). J2043+3413 was also recently studied with time-series spectra by \citet[][hereafter TH20]{2020AJ....160..151T}. The LAMOST spectrum of J2043+3413 is shown in Fig. \ref{fig:J2043spectrum}, where one can see prominent Balmer emission lines, He\uppercase\expandafter{\romannumeral2} $\lambda$4686 and C\uppercase\expandafter{\romannumeral3}/N\uppercase\expandafter{\romannumeral3} $\lambda$4650 lines. These emission lines show double-peak features, suggesting the presence of disk structure in J2043+3413. Moreover, the appearance of the prominent He \uppercase\expandafter{\romannumeral2} $\lambda$4686 emission line suggests the presence of a magnetic field (WA95) in J2043+3413, and the measured ratio of He\uppercase\expandafter{\romannumeral2} $\lambda$4686/H$\beta>0.5$ characterizes it as a candidate of NL subtype \citep{2020AJ....159...43H}. The basic information from Gaia \citep{2016A&A...595A...1G,2018A&A...616A...1G, 2022arXiv220800211G} and parameters estimated from this work are given in Table \ref{Tab:info}. 

In this paper, we examine the detailed photometric properties of J2043+3413 with our own multi-band photometry, as well as the data from TESS and ZTF. Observations, data reduction and collections of other datasets are described in Section \ref{sect:Obs}. The analysis and parameter estimations are presented in Section \ref{sect:Analysis}. Discussion and conclusion are given in Section \ref{sect:discussion}.

\begin{table*}{}

    \begin{threeparttable}
    \caption{Parameters of J2043+3413 from Gaia DR2+DR3 and this paper.}
    \begin{center}

    \setlength{\tabcolsep}{10mm}
    \begin{tabular}{l|l|l}

    \hline\hline\noalign{\smallskip}
    Parameter     &             &  Value  \\
    \hline\noalign{\smallskip}
     Right ascension \tnote{a}   &$\alpha$ [h:m:s]  & 20:43:05.955 \\[1.0ex]
     Declination \tnote{a}  &$\delta$ [d:m:s]  & +34:13:40.73 \\[1.0ex]
     Apparent magnitude \tnote{b} & $m_{\rm G}$ [mag] & 15.30 \\[1.0ex]
     BP-RP \tnote{c} &[mag] & 0.372 \\[1.0ex]
     Distance$_{G}$ \tnote{d} &$\textit{d}$ [pc]  &  $990\pm{22}$ \\ [1.0ex]
     Reddening \tnote{e} & $E(g-r)$ [mag] & $0.22_{-0.03}^{+0.02}$\\[1.0ex]
     Apparent absolute magnitude \tnote{f} &$M_{\rm G}$ [mag] & $5.32_{-0.06}^{+0.05}$\\[1.0ex]
     Orbital period \tnote{g, h} & $P_{\rm orb}$ [h] & $2.587\pm{0.008}$\\[1.0ex]
     Superorbital period \tnote{g, h} & $P_{\rm sp} (P_{\rm pr})$ [d] & $1.09\pm{0.05}$\\[1.0ex]

    \hline\noalign{\smallskip}

    \end{tabular}
    \label{Tab:info}
    \begin{tablenotes}

    \item{\textbf{Notes.}  \item[a] Positions are referred to the International Coordinate Reference System (ICRS; essentially the reference frame for J2000). \item[b]  $m_{G}$ is the mean apparent magnitude from Gaia \textit{G}-band. \item[c] The Blue (BP) and Red (RP) prism photometers collect low resolution spectrophotometric measurements of source spectral energy distribution over the wavelength ranges 330$\sim$680 nm and 630$\sim$1050 nm, respectively.  \item[d] The distance and its error are derived by the inverse of DR3 parallax. \item[e] The extinction was taken from the 3D reddening map \citep{Green_2019}. \item[f] The absolute \textit{G}-band mean magnitude is calculated from $m_{G}$ and Distance$_{G}$ \citep{2018A&A...616A...8A} as listed in this table. \item[g] See Sections \ref{sect:Analysis} and \ref{sect:discussion}. For the cases without range of error, the given values represent the rough estimates. \item[h] Analysis with the data in TESS sectors 14 and 15 (spanning from MJD 58682 to 58737).}
    \end{tablenotes}
    \end{center}
    \end{threeparttable}
\end{table*}

\section{Observation and Data Acquisition}
\label{sect:Obs}

\subsection{TNT Observation and Data Reduction}

\begin{figure}{}
   \centering
   \includegraphics[width=0.47\textwidth, angle=0]{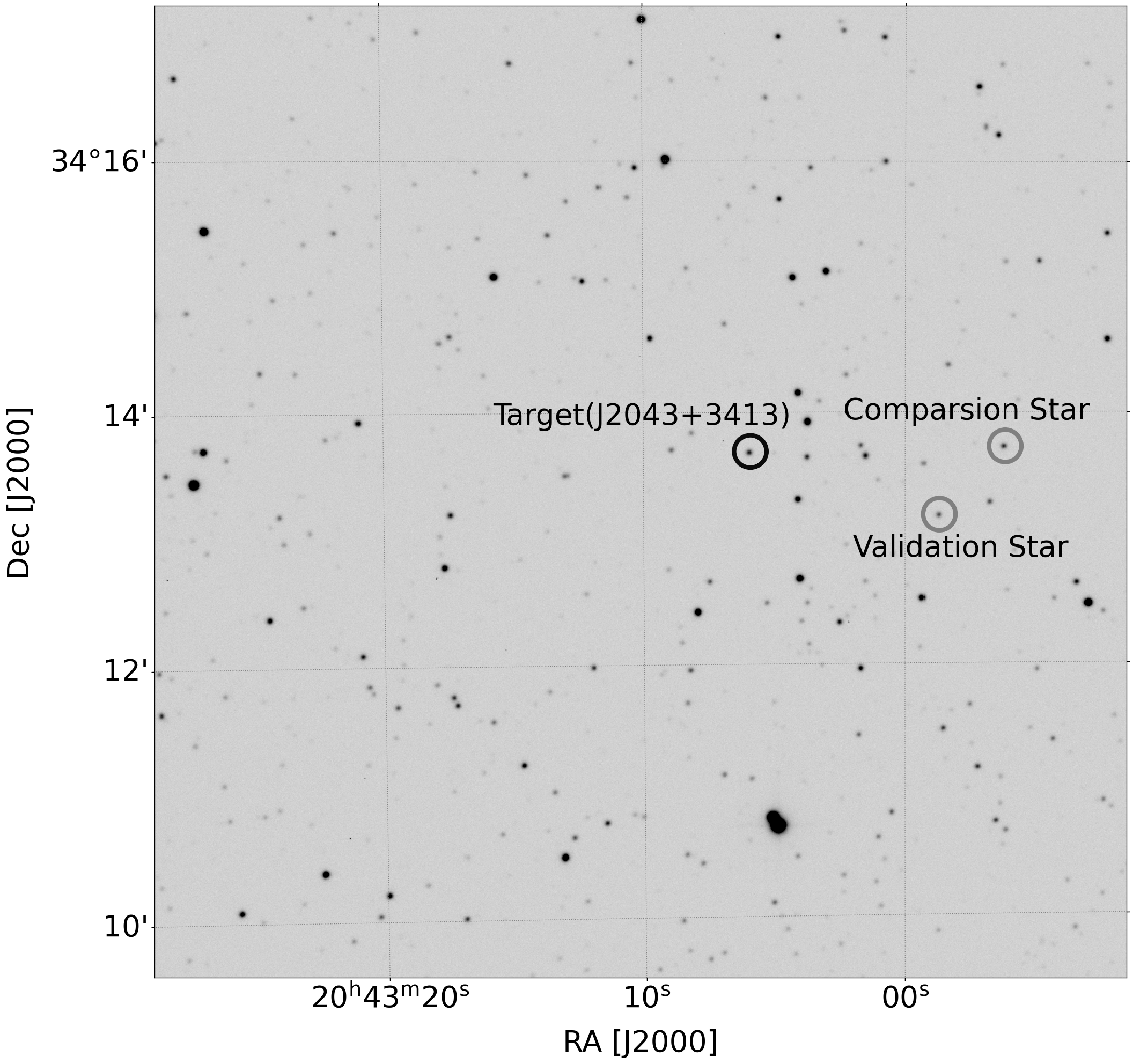}
   \caption{An observed image of the source J2043+3413. The target star and reference stars (comparison star and validation star) are marked with black and gray circles respectively.}
   \label{fig:J2043image}
\end{figure}

From October 2020 to June 2022, we observed J2043+3413 in Sloan $gri$- bands \citep{2010AJ....139.1628D} on eleven nights using the Tsinghua-NAOC 0.8-m telescope (TNT) at the Xinglong Station \citep{Wang_2008, 2012RAA....12.1585H}. The observation log of J2043+3413 is given in Table \ref{Tab:obs}. One image of this object is shown in Fig. \ref{fig:J2043image}. For long-term time-series photometric observations, the differential photometry is sufficient for our work. We select two stars with constant luminosity in the field of J2043+34 as a comparison star and validation star. All the time of our observation have been converted to Barycentric Julian Day (BJD) in Barycentric Dynamical Time (TDB) \citep{Eastman_2010}. 

\begin{table*}

\caption{Observation Journal of J2043+3413.}
\setlength{\tabcolsep}{1mm}
\label{Tab:obs}

\begin{center}
\setlength{\tabcolsep}{1.3mm}
 \begin{tabular}{lcccc}

  \hline\hline\noalign{\smallskip}

 UT Date        & Filter & Number of\ Effective Image & Exposure\ Time(s)   & Panel\ in Fig \ref{fig:J2043lc} \\                  
  \hline\noalign{\smallskip}
        2020 Oct. 24    &   \textit{i}     & 95      &80  &(a)\\ 
        2020 Oct. 25    &   \textit{r}     & 113     &80  &(b)\\
        2020 Oct. 26    &   \textit{g}     & 145     &80  &(c)\\
        2021 Jun. 18    &   \textit{r}     & 124     &80  &(d)\\
        2021 Nov. 11    &   \textit{i}     & 69      &80  &(e)\\
        2022 May. 21    &   \textit{g}     & 108     &80  &(f)\\
        2022 May. 22    &   \textit{g}     & 106     &80  &(g)\\
        2022 May. 23    &   \textit{g}     & 173     &40  &(h)\\
        2022 May. 26    &   \textit{g}     & 208     &40  &(i)\\
        2022 Jun. 05    &   \textit{g}     & 190     &40  &(j)\\
        2022 Jun. 06.   &   \textit{g}     & 191     &40  &(k)\\
        
  \noalign{\smallskip}\hline

\end{tabular}{}

\end{center}

\end{table*}

\subsection{TESS Data}

\begin{figure*}
   \centering
   \includegraphics[width=0.95\textwidth, angle=0]{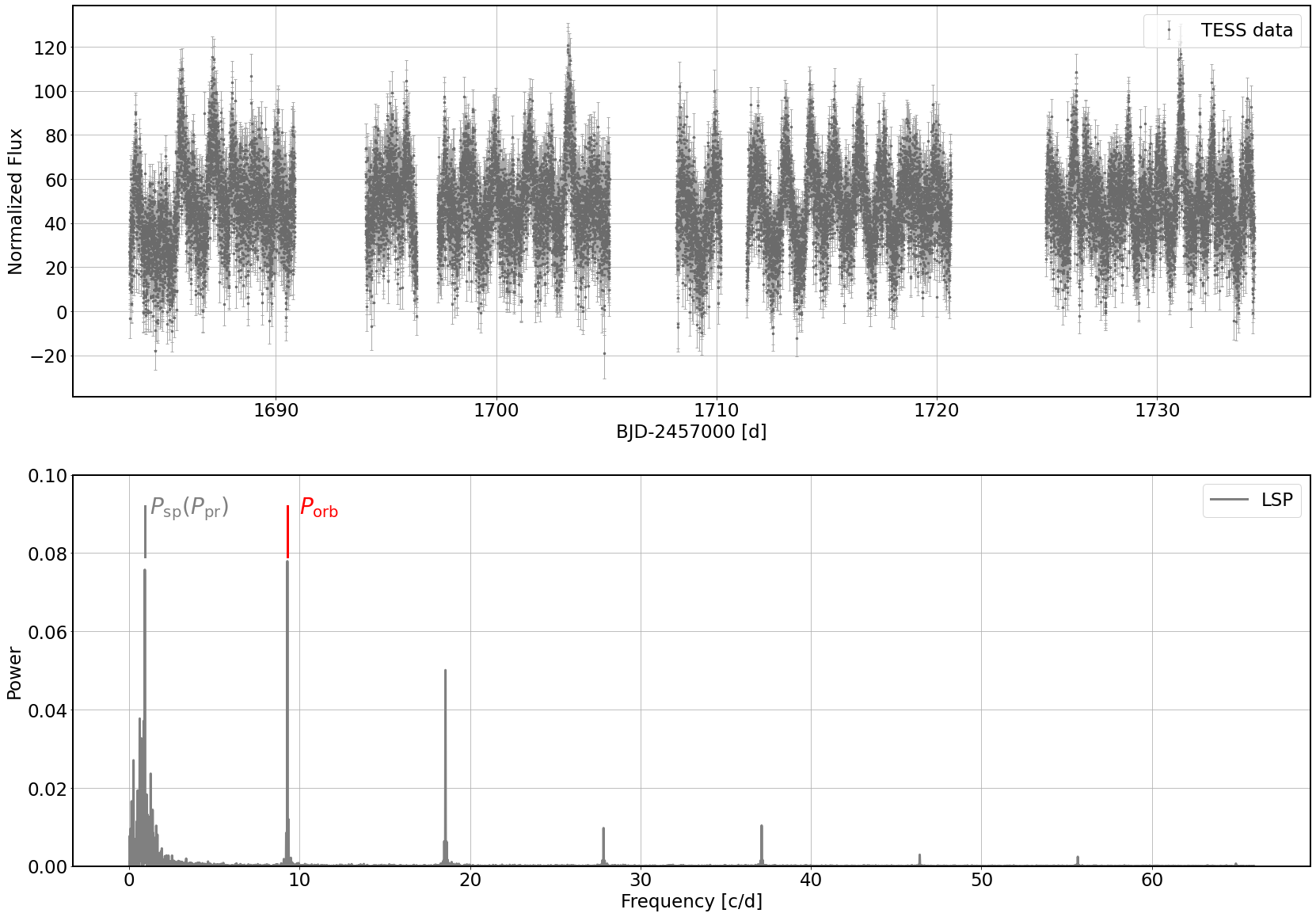}
   \caption{Upper panel: The normalized flux with the corresponding error bars (deep gray points and light gray bars) of J2043+3413 obtained within TESS 54 days monitoring. The abscissa is dated in TESS Barycentric Julian Day (BJD). Bottom panel: The LSP of J2043+3413 derived from the TESS data. The abscissa is frequency in cycles per day (c/d). The orbital period ($P_{\rm orb}$) and superorbital period ($P_{\rm sp}$) are highlighted with red and gray lines, respectively.}
   \label{fig:J2043TESSlc}
\end{figure*}

TESS works in the visible and near-infrared bands, covering wavelengths 600$\sim$1000 nm. During the first two years of its mission, TESS divided 85\% of the sky area into 26 sectors and conducted continuous observations of each sector for 27 days. For some stars located at high ecliptic latitude areas, the continuous observation can be up to 351 days due to overlapping sectors. The light curves of stars monitored by TESS are produced with MIT's Quick-Look Pipeline \citep{2020RNAAS...4..204H}. J2043+3413 (corresponding to TIC 100234005 in the TESS catalog) has been monitored by TESS in its sectors 14 and 15, and the observations lasted for 49 days with a cadence of 120s. Some of the lower quality data caused several gaps. The normalized TESS light curve of this source is shown in the upper panel of Fig. \ref{fig:J2043TESSlc}.

\subsection{Other Survey Data}

\begin{figure*}
   \centering
   \includegraphics[width=0.95\textwidth, angle=0]{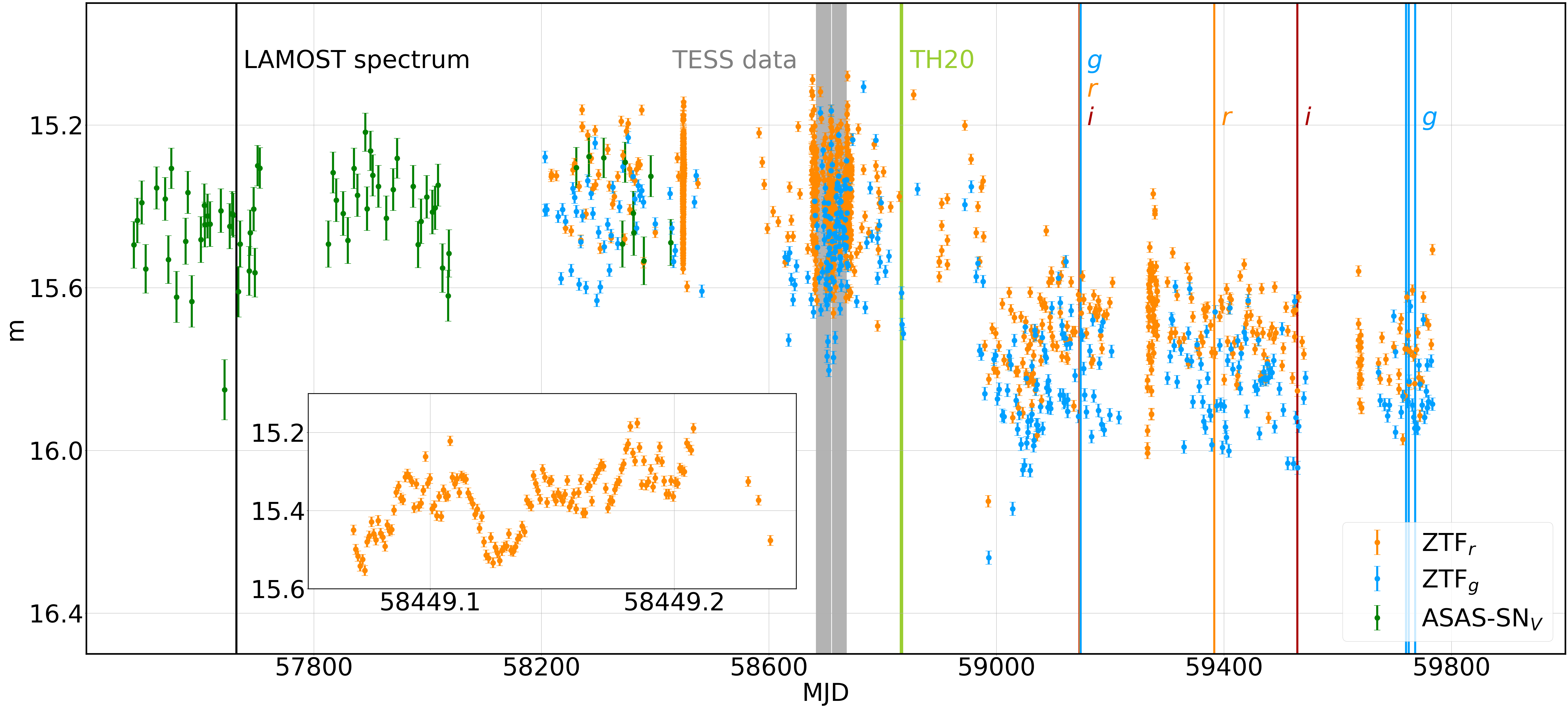}
   \caption{Observation data collected for J2043+3413. ZTF \emph{r}- and \emph{g}-bands data are shown as points in orange and blue, respectively. The only consecutive observation obtained on MJD 58449 is shown in the subgraph. Partial data of ASAS-SN \emph{V}-band is shown as points in green. The time of the TESS monitoring in which J2043+3413 was observed is marked as gray strips. The time of TH20 time-series spectroscopy is marked as yellow green line. The time of LAMOST spectrum is marked as black line. The time of TNT observation in \emph{i}-, \emph{r}- and \emph{g}-bands is marked as lines in deep red, orange and blue, respectively.}
   \label{fig:J2043ALL}
\end{figure*}

J2043+3413 has also been covered by ground-based surveys such as ZTF and ASAS-SN, which provide long-term ZTF \emph{g}/\emph{r}-bands and ASAS-SN \emph{V}/\emph{g}-bands data. Notice that both the ZTF and ASAS-SN data are under-sampled, and only one data point is available for most sidereal days. The ZTF data lasted for more than 1200 days, and the corresponding light curves of J2043+3413 are shown in Fig. \ref{fig:J2043ALL}. Only one consecutive observation was performed on the night of MJD 58449 (see the subgraph of Fig. \ref{fig:J2043ALL}).

The observation by ASAS-SN lasted for about 2000 days. Some of the $V$-band data from ASAS-SN are overplotted in Fig. \ref{fig:J2043ALL}. As the photometric errors of ASAS-SN data are relatively large, we use the ASAS-SN data only for qualitative analysis.

\section{Photometric analysis}
\label{sect:Analysis}

As can be seen from the ZTF and ASAS-SN data in Fig. \ref{fig:J2043ALL}, J2043+3413 shows a sudden decline around MJD 58979. The mean brightness of J2043+3413 is 15.47 mag in ZTF \emph{g}-band and 15.37 mag in ZTF \emph{r}-band before MJD 58979, but after that, it declined to 15.84 mag in \emph{g}-band and 15.71 mag in \emph{r}-band. Its color index also changed from 0.10 mag to 0.13 mag after MJD 58979.

\subsection{Orbital and superorbital period}
\label{sect:OP}

Relatively few CVs have been found in the orbital period range of 2$\sim$3 hours (h), the so-called CV period gap \citep{1988QJRAS..29....1K, 1998MNRAS.298L..29K}. The analysis of TH20 revealed two possible periods for J2043+3413, i.e., 2.586(3) and 2.899(3) h based on the velocities of H$\alpha$ emission line, and the later one was attributed to a daily cycle-count alias. Both of these two periods lie in the CV period gap.

We search for periodic signals using the python package {\sc Astropy} \citep{2018AJ....156..123A}. We normalized all of the light curves and computed their Lomb-Scargle periodogram (LSP here after, \citealt{1976Ap&SS..39..447L,1982ApJ...263..835S}). The LSP for the TESS data is shown in the bottom panel of Fig. \ref{fig:J2043TESSlc}.

\begin{figure}[H]
\centering
\includegraphics[width=0.47\textwidth, angle=0]{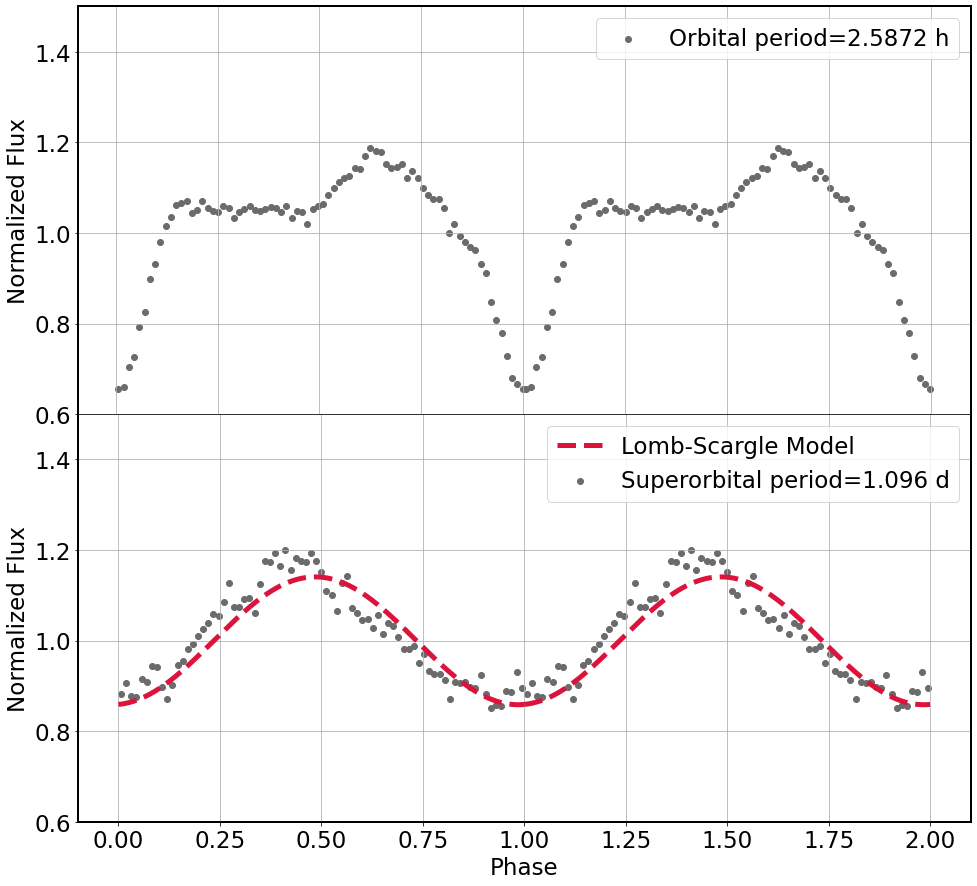}
\caption{The phase averaged normalized flux folded by the orbital period (upper panel) and superorbital period (bottom panel). The red sinusoidal represents the Lomb-Scargle model which fits well with the profile of superorbital period.}
\label{fig:J2043TESSFD}
\end{figure}

As can be seen, the LSP of the TESS data reveals several prominent peaks, the most significant ones are around 0.9127 c/d (1.096 d) and 9.2764 c/d (2.5872 h). The less significant peaks above 10 c/d are the harmonics of the fundamental frequency of 9.2764 c/d. In order to show the orbital variability, we folded the data with the period 2.5872 h, and the 120s bin-averaged normalized flux is presented in the upper panel of Fig. \ref{fig:J2043TESSFD}. The orbital light curve shows an eclipse and a pre-eclipse hump. 

To estimate the error of the orbital period, we follow the method from \citet{2018ApJS..236...16V}. The scaling of frequency is approximated by 
\begin{equation}
{\sigma_f \approx f_{1/2} \sqrt{P_{\rm max}}}
\label{equ0}
\end{equation}
where $f_{1/2}$ is the half-width of peak at half-maximum and $P_{\rm max}$ the height of peak. We obtain an error of the orbital period as 0.008 h. The orbital period derived from TESS data is $P_{\rm orb}=2.587(8)$ h, which is the same, within the uncertainties, as the 2.586(3) h period derived by TH20. 

The day-scale LSP peak around 0.9127 c/d (1.096 d) is likely a superorbital period ($P_{\rm sp}$). We estimated an error of 0.05 d. The 1200 s bin-averaged normalized flux, which is folded with superorbital periods, is shown in the bottom panel of Fig. \ref{fig:J2043TESSFD}. It can be well fitted with a sine curve (Lomb-Scargle model \citep{2018ApJS..236...16V}).

\begin{figure*}
\centering
\includegraphics[width=0.99\textwidth, angle=0]{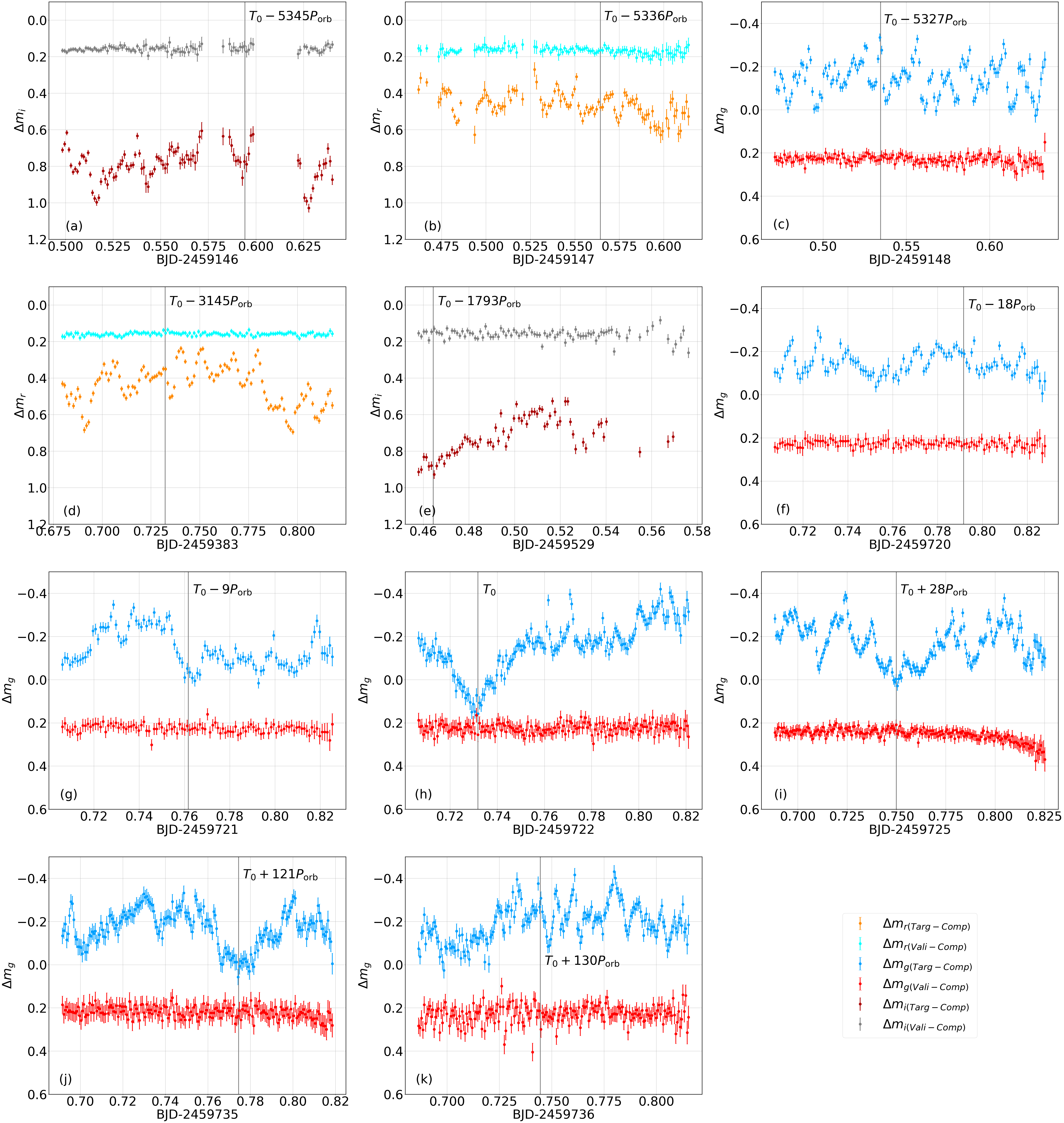}
\caption{The differential magnitudes of J2043+3413 and the reference stars, obtained by TNT on eleven nights, are plotted on the abscissa of BJD minus the observation date. The colors of the error bars corresponding to different bands and  differential magnitudes are shown in the labels. The predicted eclipse moments inferred from an eclipse-like feature ($T_0=2459722.73184$, marked in panel(h)) and the orbital period of TESS data are marked with gray lines.} 
\label{fig:J2043lc}
\end{figure*}

\subsection{Photometric variations of TNT data}

The light curves of TNT data observed on eleven nights with error bars are shown in Fig. \ref{fig:J2043lc}. It is interesting to note that the light curves on different nights are quite different. It could be seen that the light curve on 2022 May 22 (BJD 2459722) shows an eclipse-like feature, with a duration about 0.7 h, similar to the eclipsing duration of the TESS orbital profile (see the upper panel of Fig. \ref{fig:J2043TESSFD}). Based on the eclipse-like feature, we define an ephemeris of eclipse as $2459722.73184 + 0.1078 {\rm E}$, where 0.1078 d is the orbital period ($P_{\rm orb}$) obtained from TESS data. The predicted eclipse times of the other nights are plotted as vertical lines in Fig. \ref{fig:J2043lc}. $T_0$ is the time of an eclipse-like feature, which is BJD 2459722.73184. The predicted eclipse times seem to coincide with the minimum of the light curve on $T_0-9P_{\rm orb}$, $T_0+28P_{\rm orb}$, and $T_0+121P_{\rm orb}$, but not on $T_0+130P_{\rm orb}$.

\begin{figure}[H]
\centering
\includegraphics[width=0.47\textwidth, angle=0]{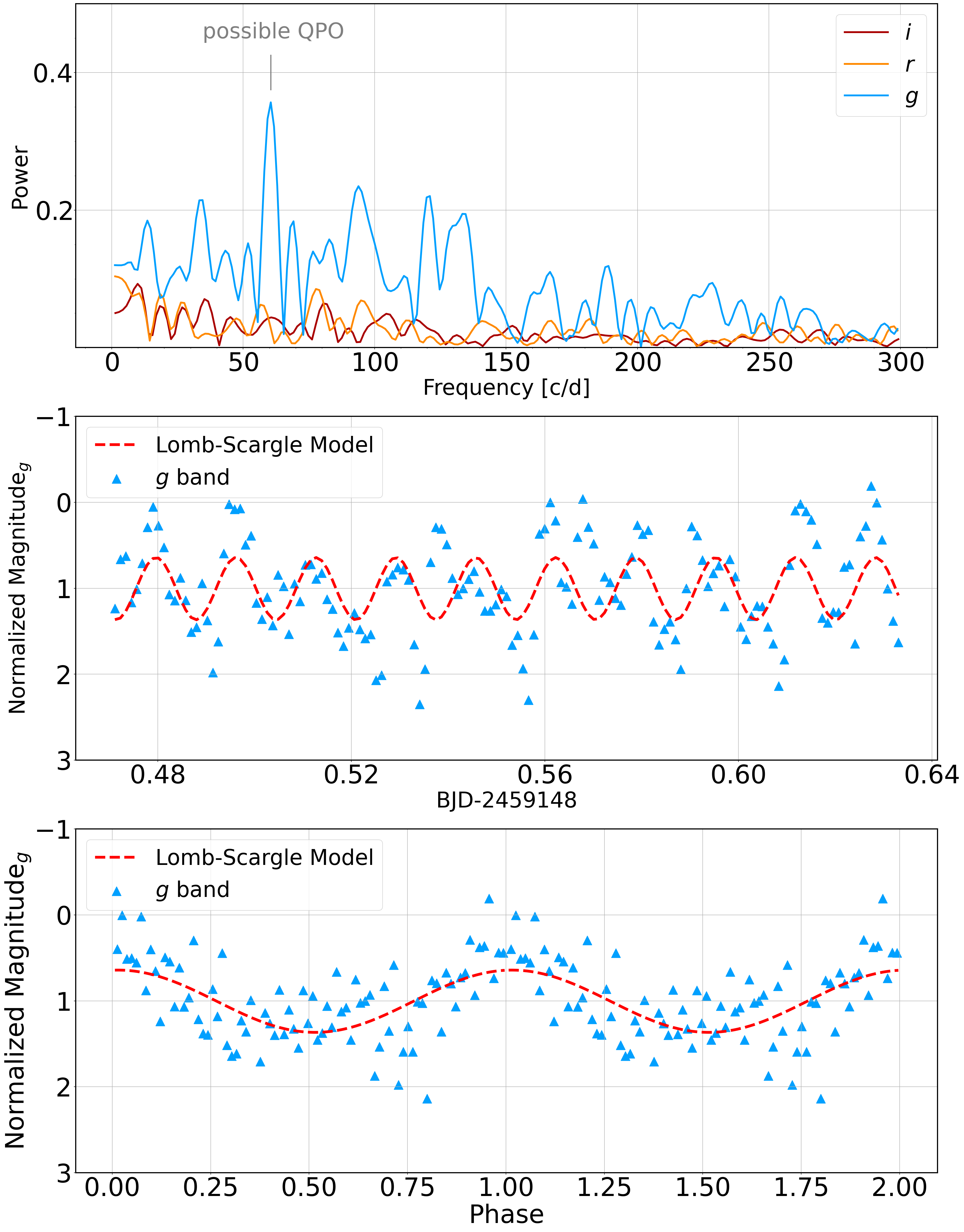}
\caption{Top panel: the LSP of J2043+34 from the TNT data observed in 2020. The possible QPO frequency $\sim$ 60.52 cycles per day is marked in gray line. Middle panel: the observed data on BJD 2459148 in \emph{g}-band. The red sinusoid represents the Lomb-Scargle model fitting. Bottom panel: The observed data folded by the fitted frequency. The time interval for the data binning is 17s.} 
\label{fig:J2043TNTLSP}
\end{figure}

The top panel of Fig. \ref{fig:J2043TNTLSP} shows the LSP of the TNT observations from 2020. We see that the LSP of \emph{i}- and \emph{r}-bands data are similar, while that of the \emph{g}-band data looks quite different. The g-band LSP shows a feature around period $P \sim$ 1426 s (60.52 c/d), which might be a kilosecond quasi-periodic oscillation (QPO; \citealt{2002MNRAS.333..411W}). The folded g-band light curve with 1426 s is presented in the bottom panel of Fig. \ref{fig:J2043TNTLSP}, with the g-band data in the middle panel. The folded light curve looks like a sine function, as represented by the Lomb-Scargle model in the figure. We also checked the LSP of TNT data on other nights and found no apparent feature.

\section{Discussion and conclusion}
\label{sect:discussion}

We have analyzed the photometric data of J0243, including TESS and ZTF survey data, as well as our own TNT data. The LSP of TESS data shows two prominent peaks (2.587 h and 1.09 d), corresponding to the orbital and super orbital periods, respectively. The superorbital period is generally attributed to the precession period of the accretion disk ($P_{\rm pr}$) in CV system \citep{1979AJ.....84..804P,1989PASJ...41.1005O,2013MNRAS.435..707A}. The observed ${P_{\rm pr}}/{P_{\rm orb}}$ of J2043+3413 is approximately 10. The orbital profile of the TESS data shows an eclipse and a pre-eclipse bump. Some of the TNT data also show eclipse-like features, while others do not.

\begin{figure}[H]
\centering
\includegraphics[width=0.47\textwidth, angle=0]{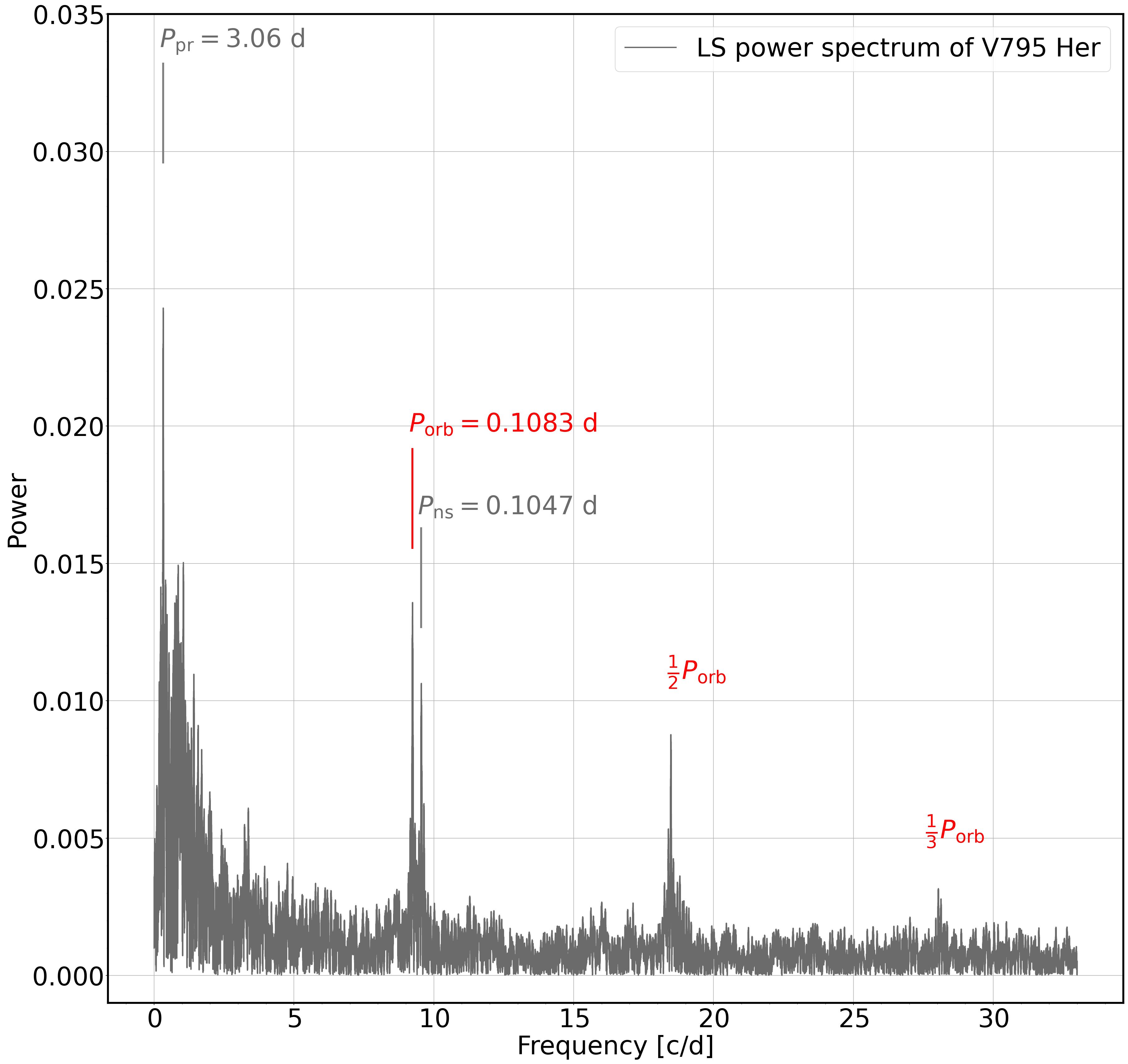}
\caption{The LSP of V795 Her derived from the TESS data, similar to the bottom panel of Fig. \ref{fig:J2043TESSlc}. The orbital period and harmonics are marked in red line. The  disk precession period and negative superhump period ($P_{\rm ns}$) \citep{2009MNRAS.398.2110W, 2021MNRAS.503.4050I} are highlighted with gray lines.}
\label{fig:V795TESSLSP}
\end{figure}

As pointed out by TH20, J2043+3413 is likely a SW Sex type NL CV, similar to V795 Her, which has an orbital period of 2.597 h. It is interesting to note that both J2043+3413 and V795 Her are located in the period gap. The previously reported ${P_{\rm pr}}/{P_{\rm orb}}$ of V795 Her is about 14, with a super orbital period of 1.53 d \citep{1990ApJ...354..708S}. We examined the TESS data of V795 Her (TIC 9464138) and found a prominent peak on 3.06 d (Fig. \ref{fig:V795TESSLSP}), which is twice the previous value. 

\begin{figure}[H]
\centering
\includegraphics[width=0.47\textwidth, angle=0]{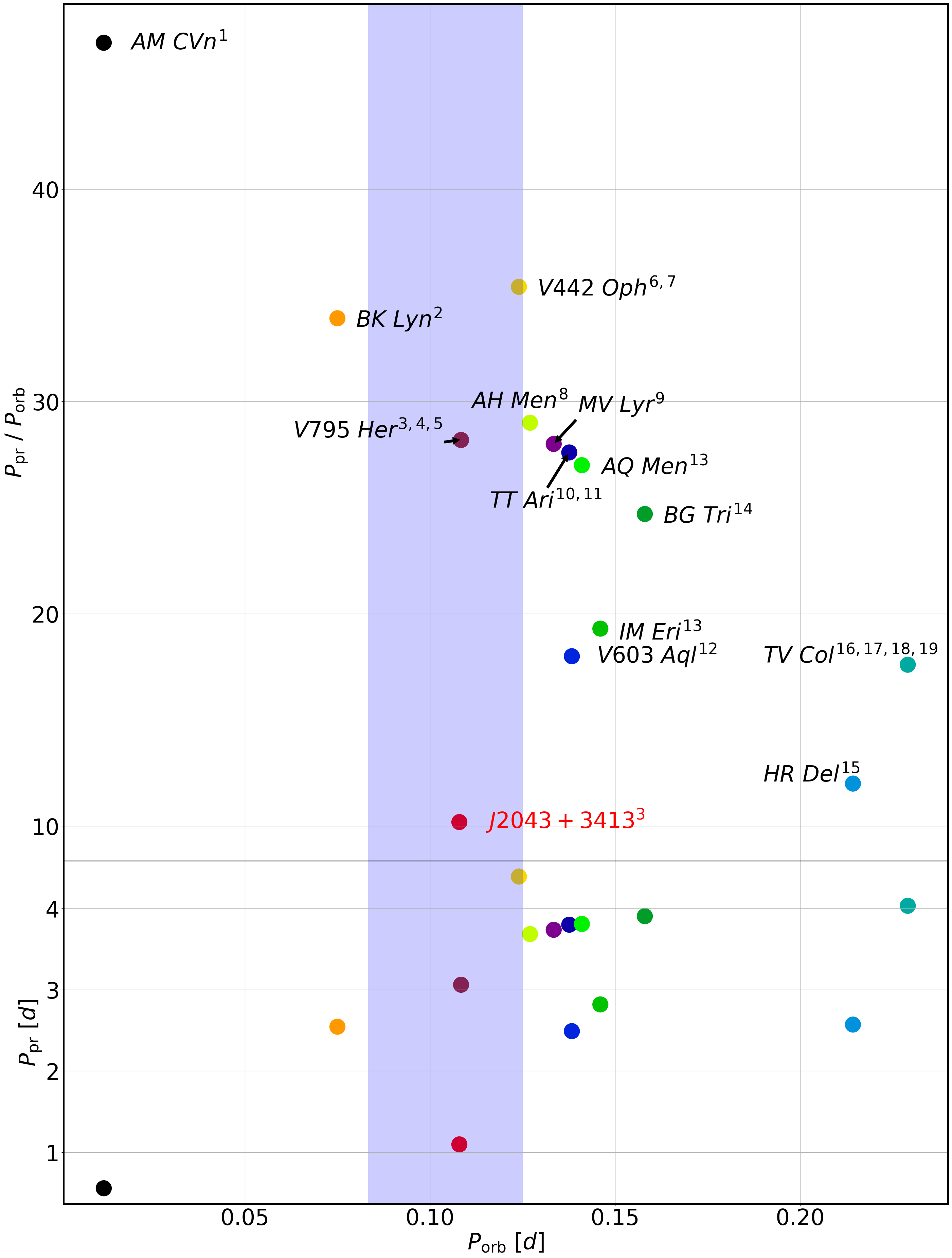}
\caption{The $P_{\rm orb}$ vs ${P_{\rm pr}}/{P_{\rm orb}}$ (top) and $P_{\rm orb}$ vs ${P_{\rm pr}}$ (bottom) for some well studied CVs, including J2043+3413 (red) and V795 Her (brown). The blue shaded area represents the 2-3 h orbital period gap of CVs. References: (1) \citet{1993ApJ...419..803P}; (2) \citet{2013MNRAS.434.1902P}; (3) this paper; (4) \citet{1990ApJ...354..708S}; (5) \citet{1996MNRAS.278..219C}; (6) \citet{2002PASP..114.1364P}; (7) \citet{2000ApJ...537..936H}; (8) \citet{1995PASP..107..657P}; (9) \citet{1992A&A...261..154B}; (10) \citet{2002ApJ...569..418W}; (11) \citet{2019MNRAS.489.2961B}; (12) \citet{1989AcA....39..125U}; (13) \citet{2013MNRAS.435..707A}; (14) \citet{2022MNRAS.tmp.2239S}; (15) \citet{1982PASP...94..916B}; (16) \citet{1994A&AS..107..219A}; (17) \citet{1985A&A...143..313B}; (18) \citet{1988MNRAS.233..759B}; (19) \citet{1993MNRAS.264..132H}.}
\label{fig:PrOrb}
\end{figure}

In Fig. \ref{fig:PrOrb} we compare the ${P_{\rm pr}}/{P_{\rm orb}}$ of J2043+3413 with some well-studied CVs. Besides V795 Her, TT Ari, AH Men and V442 Oph also belong to SW Sex stars. IM Eri, AQ Men and MV Lyr are classified as other NL subclasses. AM CVn is a hydrogen-deficient CV with a short orbital period. V603 Aql and HR Del belong to the CN subclass. BK Lyn is a confirmed DN system. TV Col is a long-period IP. The ${P_{\rm pr}}/{P_{\rm orb}}$ of J2043+3413 is 2-3 times less than the general trend of ${P_{\rm pr}}/{P_{\rm orb}}$, showing that J2043+3413 is a very special CV. The precession period of J2043+3413 is 2$\sim$3 times shorter than those of others.

QPOs detected in CVs are most likely from the accretion disks \citep{1985ApJ...296..529C, 1993ApJ...409..360L}. Vertical or radial oscillations of accretion disks have been proposed as the probable origin of QPOs. Some QPOs may be related to magnetically-controlled accretion in magnetic CVs (WA95). Long period (kilosecond) QPOs are even more confusing phenomena, which have been observed in some SW Sex stars \citep{Lima_2021}. They may reflect the spin of primary WD, which is drowned by the noise of high accretion rate in these systems \citep{2002PASP..114.1364P, 2003AJ....126.2473H}.

The LSP of \emph{g}-band TNT data observed in 2020 shows a peak around 1426 s, which may be a transient QPO phenomenon. It is interesting to note that V795 Her was also found to show a transient QPO around 1160 s \citep{1994PASP..106.1141P}. The intermittent behavior of the eclipse of J2043+3413 indicates an intermediate orbital inclination. This is similar to V795 Her, whose inclination is $i\approx56^\circ$ \citep{1996MNRAS.278..219C}.

In summary, J2043+3413 is a SW Sex candidate with an orbital period, an intermediate inclination, and a transient QPO, very similar to V795 Her. But the precession period of J2043+3413 is about 3 times shorter compared to V795 Her and other typical CVs, indicating an unusual, fast precessing disk in J2043+3413. The physical mechanism of the faster precession of J2043+3413 is unclear to us and further studies are needed to reveal their nature.

\section*{Acknowledgments}

We thank our referee for very detailed and great suggestions, which help us a lot to improve this paper. We thank Fengwu Sun from University of Arizona for helpful suggestions. This work is supported by the National Natural Science Foundation of China (NSFC grants 12288102, 12033003, 11633002 and 12203006), Science Program of Beijing Academy of Science and Technology (23CB061, BS202002), and the Tencent Xplorer Prize. We acknowledge the support of the staff of the Xinglong 80cm telescope (TNT).

The TESS data presented in this paper were obtained from the Mikulski Archive for Space Telescopes (MAST) at the Space Telescope Science Institute (STScI). The specific observations analyzed can be accessed via \dataset[10.17909/t9-nmc8-f686]{https://doi.org/10.17909/t9-nmc8-f686}. This work has made use of data from the European Space Agency (ESA) mission Gaia (\href{https://www.cosmos.esa.int/gaia}{https://www.cosmos.esa.int/gaia}), processed by the Gaia Data Processing and Analysis Consortium (DPAC, \href{https:// www.cosmos.esa.int/web/gaia/dpac/consortium}{https:// www.cosmos.esa.int/web/gaia/dpac/consortium}). Funding for the DPAC has been provided by national institutions, in particular the institutions participating in the Gaia Multilateral Agreement. Based on observations obtained with the Samuel Oschin Telescope 48 inch and the 60 inch Telescope at the Palomar Observatory as part of the 
ZTF project. ZTF is supported by the NSF under grant AST-1440341 and a collaboration including Caltech, IPAC, the Weizmann Institute for Science, the Oskar Klein Center at Stockholm University, the University of Maryland, the University of Washington, Deutsches Elektronen-Synchrotron and Humboldt University, Los Alamos National Laboratories, the TANGO Consortium of Taiwan, the University of Wisconsin at Milwaukee, and Lawrence Berkeley national Laboratories. Funding for LAMOST (\href{www.lamost.org}{http://www.lamost.org}) has been provided by the Chinese NDRC. LAMOST is operated and managed by the NAOC.

\bibliography{J2043+3413}{}

\begin{thebibliography}{}
\expandafter\ifx\csname natexlab\endcsname\relax\def\natexlab#1{#1}\fi
\providecommand{\url}[1]{\href{#1}{#1}}
\providecommand{\dodoi}[1]{doi:~\href{http://doi.org/#1}{\nolinkurl{#1}}}
\providecommand{\doeprint}[1]{\href{http://ascl.net/#1}{\nolinkurl{http://ascl.net/#1}}}
\providecommand{\doarXiv}[1]{\href{https://arxiv.org/abs/#1}{\nolinkurl{https://arxiv.org/abs/#1}}}

\bibitem[{{Andrae} {et~al.}(2018){Andrae}, {Fouesneau}, {Creevey}, {Ordenovic},
  {Mary}, {Burlacu}, {Chaoul}, {Jean-Antoine-Piccolo}, {Kordopatis}, {Korn},
  {Lebreton}, {Panem}, {Pichon}, {Th{\'e}venin}, {Walmsley}, \&
  {Bailer-Jones}}]{2018A&A...616A...8A}
{Andrae}, R., {Fouesneau}, M., {Creevey}, O., {et~al.} 2018, \aap, 616, A8,
  \dodoi{10.1051/0004-6361/201732516}

\bibitem[{{Armstrong} {et~al.}(2013){Armstrong}, {Patterson}, {Michelsen},
  {Thorstensen}, {Uthas}, {Vanmunster}, {Hambsch}, {Roberts}, \&
  {Dvorak}}]{2013MNRAS.435..707A}
{Armstrong}, E., {Patterson}, J., {Michelsen}, E., {et~al.} 2013, \mnras, 435,
  707, \dodoi{10.1093/mnras/stt1335}

\bibitem[{{Astropy Collaboration} {et~al.}(2018){Astropy Collaboration},
  {Price-Whelan}, {Sip{\H{o}}cz}, {G{\"u}nther}, {Lim}, {Crawford}, {Conseil},
  {Shupe}, {Craig}, {Dencheva}, {Ginsburg}, {VanderPlas}, {Bradley},
  {P{\'e}rez-Su{\'a}rez}, {de Val-Borro}, {Aldcroft}, {Cruz}, {Robitaille},
  {Tollerud}, {Ardelean}, {Babej}, {Bach}, {Bachetti}, {Bakanov}, {Bamford},
  {Barentsen}, {Barmby}, {Baumbach}, {Berry}, {Biscani}, {Boquien}, {Bostroem},
  {Bouma}, {Brammer}, {Bray}, {Breytenbach}, {Buddelmeijer}, {Burke},
  {Calderone}, {Cano Rodr{\'\i}guez}, {Cara}, {Cardoso}, {Cheedella}, {Copin},
  {Corrales}, {Crichton}, {D'Avella}, {Deil}, {Depagne}, {Dietrich}, {Donath},
  {Droettboom}, {Earl}, {Erben}, {Fabbro}, {Ferreira}, {Finethy}, {Fox},
  {Garrison}, {Gibbons}, {Goldstein}, {Gommers}, {Greco}, {Greenfield},
  {Groener}, {Grollier}, {Hagen}, {Hirst}, {Homeier}, {Horton}, {Hosseinzadeh},
  {Hu}, {Hunkeler}, {Ivezi{\'c}}, {Jain}, {Jenness}, {Kanarek}, {Kendrew},
  {Kern}, {Kerzendorf}, {Khvalko}, {King}, {Kirkby}, {Kulkarni}, {Kumar},
  {Lee}, {Lenz}, {Littlefair}, {Ma}, {Macleod}, {Mastropietro}, {McCully},
  {Montagnac}, {Morris}, {Mueller}, {Mumford}, {Muna}, {Murphy}, {Nelson},
  {Nguyen}, {Ninan}, {N{\"o}the}, {Ogaz}, {Oh}, {Parejko}, {Parley}, {Pascual},
  {Patil}, {Patil}, {Plunkett}, {Prochaska}, {Rastogi}, {Reddy Janga},
  {Sabater}, {Sakurikar}, {Seifert}, {Sherbert}, {Sherwood-Taylor}, {Shih},
  {Sick}, {Silbiger}, {Singanamalla}, {Singer}, {Sladen}, {Sooley},
  {Sornarajah}, {Streicher}, {Teuben}, {Thomas}, {Tremblay}, {Turner},
  {Terr{\'o}n}, {van Kerkwijk}, {de la Vega}, {Watkins}, {Weaver}, {Whitmore},
  {Woillez}, {Zabalza}, \& {Astropy Contributors}}]{2018AJ....156..123A}
{Astropy Collaboration}, {Price-Whelan}, A.~M., {Sip{\H{o}}cz}, B.~M., {et~al.}
  2018, \aj, 156, 123, \dodoi{10.3847/1538-3881/aabc4f}

\bibitem[{{Augusteijn} {et~al.}(1994){Augusteijn}, {Heemskerk}, {Zwarthoed}, \&
  {van Paradijs}}]{1994A&AS..107..219A}
{Augusteijn}, T., {Heemskerk}, M.~H.~M., {Zwarthoed}, G.~A.~A., \& {van
  Paradijs}, J. 1994, \aaps, 107, 219

\bibitem[{{Barrett} {et~al.}(1988){Barrett}, {O'Donoghue}, \&
  {Warner}}]{1988MNRAS.233..759B}
{Barrett}, P., {O'Donoghue}, D., \& {Warner}, B. 1988, \mnras, 233, 759,
  \dodoi{10.1093/mnras/233.4.759}

\bibitem[{Bellm {et~al.}(2019)Bellm, Kulkarni, Barlow, Feindt, Graham, Goobar,
  Kupfer, Ngeow, Nugent, Ofek, Prince, Riddle, Walters, \& Ye}]{Bellm_2019}
Bellm, E.~C., Kulkarni, S.~R., Barlow, T., {et~al.} 2019, \pasp, 131, 068003,
  \dodoi{10.1088/1538-3873/ab0c2a}

\bibitem[{{Bonnet-Bidaud} {et~al.}(1985){Bonnet-Bidaud}, {Motch}, \&
  {Mouchet}}]{1985A&A...143..313B}
{Bonnet-Bidaud}, J.~M., {Motch}, C., \& {Mouchet}, M. 1985, \aap, 143, 313

\bibitem[{{Borisov}(1992)}]{1992A&A...261..154B}
{Borisov}, G.~V. 1992, \aap, 261, 154

\bibitem[{{Bruch}(1982)}]{1982PASP...94..916B}
{Bruch}, A. 1982, \pasp, 94, 916, \dodoi{10.1086/131085}

\bibitem[{{Bruch}(2019)}]{2019MNRAS.489.2961B}
---. 2019, \mnras, 489, 2961, \dodoi{10.1093/mnras/stz2381}

\bibitem[{{Carroll} {et~al.}(1985){Carroll}, {McDermott}, {Savedoff}, {van
  Horn}, \& {Cabot}}]{1985ApJ...296..529C}
{Carroll}, B.~W., {McDermott}, P.~N., {Savedoff}, M.~P., {van Horn}, H.~M., \&
  {Cabot}, W. 1985, \apj, 296, 529, \dodoi{10.1086/163472}

\bibitem[{{Casares} {et~al.}(1996){Casares}, {Martinez-Pais}, {Marsh},
  {Charles}, \& {Lazaro}}]{1996MNRAS.278..219C}
{Casares}, J., {Martinez-Pais}, I.~G., {Marsh}, T.~R., {Charles}, P.~A., \&
  {Lazaro}, C. 1996, \mnras, 278, 219, \dodoi{10.1093/mnras/278.1.219}

\bibitem[{{Cropper}(1990)}]{1990SSRv...54..195C}
{Cropper}, M. 1990, \ssr, 54, 195, \dodoi{10.1007/BF00177799}

\bibitem[{{Cropper} {et~al.}(2002){Cropper}, {Ramsay}, {Hellier}, {Mukai},
  {Mauche}, \& {Pandel}}]{2002RSPTA.360.1951C}
{Cropper}, M., {Ramsay}, G., {Hellier}, C., {et~al.} 2002, Philosophical
  Transactions of the Royal Society of London Series A, 360, 1951,
  \dodoi{10.1098/rsta.2002.1046}

\bibitem[{{Cui} {et~al.}(2012){Cui}, {Zhao}, {Chu}, {Li}, {Li}, {Zhang}, {Su},
  {Yao}, {Wang}, {Xing}, {Li}, {Zhu}, {Wang}, {Gu}, {Luo}, {Xu}, {Zhang},
  {Liu}, {Zhang}, {Yang}, {Cao}, {Chen}, {Chen}, {Chen}, {Chen}, {Chu}, {Feng},
  {Gong}, {Hou}, {Hu}, {Hu}, {Hu}, {Jia}, {Jiang}, {Jiang}, {Jiang}, {Jin},
  {Li}, {Li}, {Li}, {Liu}, {Liu}, {Lu}, {Mao}, {Men}, {Qi}, {Qi}, {Shi},
  {Tang}, {Tao}, {Wang}, {Wang}, {Wang}, {Wang}, {Wang}, {Wang}, {Wang},
  {Wang}, {Wang}, {Wang}, {Wang}, {Wang}, {Xu}, {Xu}, {Yang}, {Yu}, {Yuan},
  {Yuan}, {Zhai}, {Zhang}, {Zhang}, {Zhang}, {Zhao}, {Zhou}, {Zhou}, {Zhu}, \&
  {Zou}}]{2012RAA....12.1197C}
{Cui}, X.-Q., {Zhao}, Y.-H., {Chu}, Y.-Q., {et~al.} 2012, Research in Astronomy
  and Astrophysics, 12, 1197, \dodoi{10.1088/1674-4527/12/9/003}

\bibitem[{Dai {et~al.}(2016)Dai, Szkody, Garnavich, \& Kennedy}]{Dai_2016}
Dai, Z., Szkody, P., Garnavich, P.~M., \& Kennedy, M. 2016, The Astronomical
  Journal, 152, 5, \dodoi{10.3847/0004-6256/152/1/5}

\bibitem[{{Doi} {et~al.}(2010){Doi}, {Tanaka}, {Fukugita}, {Gunn}, {Yasuda},
  {Ivezi{\'c}}, {Brinkmann}, {de Haars}, {Kleinman}, {Krzesinski}, \& {French
  Leger}}]{2010AJ....139.1628D}
{Doi}, M., {Tanaka}, M., {Fukugita}, M., {et~al.} 2010, \aj, 139, 1628,
  \dodoi{10.1088/0004-6256/139/4/1628}

\bibitem[{Eastman {et~al.}(2010)Eastman, Siverd, \& Gaudi}]{Eastman_2010}
Eastman, J., Siverd, R., \& Gaudi, B.~S. 2010, Publications of the Astronomical
  Society of the Pacific, 122, 935, \dodoi{10.1086/655938}

\bibitem[{{Gaia Collaboration} {et~al.}(2016){Gaia Collaboration}, {Prusti},
  {de Bruijne}, {Brown}, {Vallenari}, {Babusiaux}, {Bailer-Jones}, {Bastian},
  {Biermann}, {Evans}, {Eyer}, {Jansen}, {Jordi}, {Klioner}, {Lammers},
  {Lindegren}, {Luri}, {Mignard}, {Milligan}, {Panem}, {Poinsignon},
  {Pourbaix}, {Randich}, {Sarri}, {Sartoretti}, {Siddiqui}, {Soubiran},
  {Valette}, {van Leeuwen}, {Walton}, {Aerts}, {Arenou}, {Cropper}, {Drimmel},
  {H{\o}g}, {Katz}, {Lattanzi}, {O'Mullane}, {Grebel}, {Holland}, {Huc},
  {Passot}, {Bramante}, {Cacciari}, {Casta{\~n}eda}, {Chaoul}, {Cheek}, {De
  Angeli}, {Fabricius}, {Guerra}, {Hern{\'a}ndez}, {Jean-Antoine-Piccolo},
  {Masana}, {Messineo}, {Mowlavi}, {Nienartowicz}, {Ord{\'o}{\~n}ez-Blanco},
  {Panuzzo}, {Portell}, {Richards}, {Riello}, {Seabroke}, {Tanga},
  {Th{\'e}venin}, {Torra}, {Els}, {Gracia-Abril}, {Comoretto},
  {Garcia-Reinaldos}, {Lock}, {Mercier}, {Altmann}, {Andrae}, {Astraatmadja},
  {Bellas-Velidis}, {Benson}, {Berthier}, {Blomme}, {Busso}, {Carry},
  {Cellino}, {Clementini}, {Cowell}, {Creevey}, {Cuypers}, {Davidson}, {De
  Ridder}, {de Torres}, {Delchambre}, {Dell'Oro}, {Ducourant}, {Fr{\'e}mat},
  {Garc{\'\i}a-Torres}, {Gosset}, {Halbwachs}, {Hambly}, {Harrison}, {Hauser},
  {Hestroffer}, {Hodgkin}, {Huckle}, {Hutton}, {Jasniewicz}, {Jordan},
  {Kontizas}, {Korn}, {Lanzafame}, {Manteiga}, {Moitinho}, {Muinonen},
  {Osinde}, {Pancino}, {Pauwels}, {Petit}, {Recio-Blanco}, {Robin}, {Sarro},
  {Siopis}, {Smith}, {Smith}, {Sozzetti}, {Thuillot}, {van Reeven}, {Viala},
  {Abbas}, {Abreu Aramburu}, {Accart}, {Aguado}, {Allan}, {Allasia},
  {Altavilla}, {{\'A}lvarez}, {Alves}, {Anderson}, {Andrei}, {Anglada Varela},
  {Antiche}, {Antoja}, {Ant{\'o}n}, {Arcay}, {Atzei}, {Ayache}, {Bach},
  {Baker}, {Balaguer-N{\'u}{\~n}ez}, {Barache}, {Barata}, {Barbier}, {Barblan},
  {Baroni}, {Barrado y Navascu{\'e}s}, {Barros}, {Barstow}, {Becciani},
  {Bellazzini}, {Bellei}, {Bello Garc{\'\i}a}, {Belokurov}, {Bendjoya},
  {Berihuete}, {Bianchi}, {Bienaym{\'e}}, {Billebaud}, {Blagorodnova},
  {Blanco-Cuaresma}, {Boch}, {Bombrun}, {Borrachero}, {Bouquillon}, {Bourda},
  {Bouy}, {Bragaglia}, {Breddels}, {Brouillet}, {Br{\"u}semeister},
  {Bucciarelli}, {Budnik}, {Burgess}, {Burgon}, {Burlacu}, {Busonero}, {Buzzi},
  {Caffau}, {Cambras}, {Campbell}, {Cancelliere}, {Cantat-Gaudin}, {Carlucci},
  {Carrasco}, {Castellani}, {Charlot}, {Charnas}, {Charvet}, {Chassat},
  {Chiavassa}, {Clotet}, {Cocozza}, {Collins}, {Collins}, {Costigan}, {Crifo},
  {Cross}, {Crosta}, {Crowley}, {Dafonte}, {Damerdji}, {Dapergolas}, {David},
  {David}, {De Cat}, {de Felice}, {de Laverny}, {De Luise}, {De March}, {de
  Martino}, {de Souza}, {Debosscher}, {del Pozo}, {Delbo}, {Delgado},
  {Delgado}, {di Marco}, {Di Matteo}, {Diakite}, {Distefano}, {Dolding}, {Dos
  Anjos}, {Drazinos}, {Dur{\'a}n}, {Dzigan}, {Ecale}, {Edvardsson}, {Enke},
  {Erdmann}, {Escolar}, {Espina}, {Evans}, {Eynard Bontemps}, {Fabre},
  {Fabrizio}, {Faigler}, {Falc{\~a}o}, {Farr{\`a}s Casas}, {Faye}, {Federici},
  {Fedorets}, {Fern{\'a}ndez-Hern{\'a}ndez}, {Fernique}, {Fienga}, {Figueras},
  {Filippi}, {Findeisen}, {Fonti}, {Fouesneau}, {Fraile}, {Fraser}, {Fuchs},
  {Furnell}, {Gai}, {Galleti}, {Galluccio}, {Garabato}, {Garc{\'\i}a-Sedano},
  {Gar{\'e}}, {Garofalo}, {Garralda}, {Gavras}, {Gerssen}, {Geyer}, {Gilmore},
  {Girona}, {Giuffrida}, {Gomes}, {Gonz{\'a}lez-Marcos},
  {Gonz{\'a}lez-N{\'u}{\~n}ez}, {Gonz{\'a}lez-Vidal}, {Granvik}, {Guerrier},
  {Guillout}, {Guiraud}, {G{\'u}rpide}, {Guti{\'e}rrez-S{\'a}nchez}, {Guy},
  {Haigron}, {Hatzidimitriou}, {Haywood}, {Heiter}, {Helmi}, {Hobbs},
  {Hofmann}, {Holl}, {Holland}, {Hunt}, {Hypki}, {Icardi}, {Irwin}, {Jevardat
  de Fombelle}, {Jofr{\'e}}, {Jonker}, {Jorissen}, {Julbe}, {Karampelas},
  {Kochoska}, {Kohley}, {Kolenberg}, {Kontizas}, {Koposov}, {Kordopatis},
  {Koubsky}, {Kowalczyk}, {Krone-Martins}, {Kudryashova}, {Kull}, {Bachchan},
  {Lacoste-Seris}, {Lanza}, {Lavigne}, {Le Poncin-Lafitte}, {Lebreton},
  {Lebzelter}, {Leccia}, {Leclerc}, {Lecoeur-Taibi}, {Lemaitre}, {Lenhardt},
  {Leroux}, {Liao}, {Licata}, {Lindstr{\o}m}, {Lister}, {Livanou}, {Lobel},
  {L{\"o}ffler}, {L{\'o}pez}, {Lopez-Lozano}, {Lorenz}, {Loureiro},
  {MacDonald}, {Magalh{\~a}es Fernandes}, {Managau}, {Mann}, {Mantelet},
  {Marchal}, {Marchant}, {Marconi}, {Marie}, {Marinoni}, {Marrese},
  {Marschalk{\'o}}, {Marshall}, {Mart{\'\i}n-Fleitas}, {Martino}, {Mary},
  {Matijevi{\v{c}}}, {Mazeh}, {McMillan}, {Messina}, {Mestre}, {Michalik},
  {Millar}, {Miranda}, {Molina}, {Molinaro}, {Molinaro}, {Moln{\'a}r},
  {Moniez}, {Montegriffo}, {Monteiro}, {Mor}, {Mora}, {Morbidelli}, {Morel},
  {Morgenthaler}, {Morley}, {Morris}, {Mulone}, {Muraveva}, {Musella},
  {Narbonne}, {Nelemans}, {Nicastro}, {Noval}, {Ord{\'e}novic},
  {Ordieres-Mer{\'e}}, {Osborne}, {Pagani}, {Pagano}, {Pailler}, {Palacin},
  {Palaversa}, {Parsons}, {Paulsen}, {Pecoraro}, {Pedrosa}, {Pentik{\"a}inen},
  {Pereira}, {Pichon}, {Piersimoni}, {Pineau}, {Plachy}, {Plum}, {Poujoulet},
  {Pr{\v{s}}a}, {Pulone}, {Ragaini}, {Rago}, {Rambaux}, {Ramos-Lerate},
  {Ranalli}, {Rauw}, {Read}, {Regibo}, {Renk}, {Reyl{\'e}}, {Ribeiro},
  {Rimoldini}, {Ripepi}, {Riva}, {Rixon}, {Roelens}, {Romero-G{\'o}mez},
  {Rowell}, {Royer}, {Rudolph}, {Ruiz-Dern}, {Sadowski}, {Sagrist{\`a}
  Sell{\'e}s}, {Sahlmann}, {Salgado}, {Salguero}, {Sarasso}, {Savietto},
  {Schnorhk}, {Schultheis}, {Sciacca}, {Segol}, {Segovia}, {Segransan},
  {Serpell}, {Shih}, {Smareglia}, {Smart}, {Smith}, {Solano}, {Solitro},
  {Sordo}, {Soria Nieto}, {Souchay}, {Spagna}, {Spoto}, {Stampa}, {Steele},
  {Steidelm{\"u}ller}, {Stephenson}, {Stoev}, {Suess}, {S{\"u}veges}, {Surdej},
  {Szabados}, {Szegedi-Elek}, {Tapiador}, {Taris}, {Tauran}, {Taylor},
  {Teixeira}, {Terrett}, {Tingley}, {Trager}, {Turon}, {Ulla}, {Utrilla},
  {Valentini}, {van Elteren}, {Van Hemelryck}, {van Leeuwen}, {Varadi},
  {Vecchiato}, {Veljanoski}, {Via}, {Vicente}, {Vogt}, {Voss}, {Votruba},
  {Voutsinas}, {Walmsley}, {Weiler}, {Weingrill}, {Werner}, {Wevers},
  {Whitehead}, {Wyrzykowski}, {Yoldas}, {{\v{Z}}erjal}, {Zucker}, {Zurbach},
  {Zwitter}, {Alecu}, {Allen}, {Allende Prieto}, {Amorim},
  {Anglada-Escud{\'e}}, {Arsenijevic}, {Azaz}, {Balm}, {Beck}, {Bernstein},
  {Bigot}, {Bijaoui}, {Blasco}, {Bonfigli}, {Bono}, {Boudreault}, {Bressan},
  {Brown}, {Brunet}, {Bunclark}, {Buonanno}, {Butkevich}, {Carret}, {Carrion},
  {Chemin}, {Ch{\'e}reau}, {Corcione}, {Darmigny}, {de Boer}, {de Teodoro}, {de
  Zeeuw}, {Delle Luche}, {Domingues}, {Dubath}, {Fodor}, {Fr{\'e}zouls},
  {Fries}, {Fustes}, {Fyfe}, {Gallardo}, {Gallegos}, {Gardiol}, {Gebran},
  {Gomboc}, {G{\'o}mez}, {Grux}, {Gueguen}, {Heyrovsky}, {Hoar}, {Iannicola},
  {Isasi Parache}, {Janotto}, {Joliet}, {Jonckheere}, {Keil}, {Kim},
  {Klagyivik}, {Klar}, {Knude}, {Kochukhov}, {Kolka}, {Kos}, {Kutka}, {Lainey},
  {LeBouquin}, {Liu}, {Loreggia}, {Makarov}, {Marseille}, {Martayan},
  {Martinez-Rubi}, {Massart}, {Meynadier}, {Mignot}, {Munari}, {Nguyen},
  {Nordlander}, {Ocvirk}, {O'Flaherty}, {Olias Sanz}, {Ortiz}, {Osorio},
  {Oszkiewicz}, {Ouzounis}, {Palmer}, {Park}, {Pasquato}, {Peltzer}, {Peralta},
  {P{\'e}turaud}, {Pieniluoma}, {Pigozzi}, {Poels}, {Prat}, {Prod'homme},
  {Raison}, {Rebordao}, {Risquez}, {Rocca-Volmerange}, {Rosen}, {Ruiz-Fuertes},
  {Russo}, {Sembay}, {Serraller Vizcaino}, {Short}, {Siebert}, {Silva},
  {Sinachopoulos}, {Slezak}, {Soffel}, {Sosnowska}, {Strai{\v{z}}ys}, {ter
  Linden}, {Terrell}, {Theil}, {Tiede}, {Troisi}, {Tsalmantza}, {Tur},
  {Vaccari}, {Vachier}, {Valles}, {Van Hamme}, {Veltz}, {Virtanen}, {Wallut},
  {Wichmann}, {Wilkinson}, {Ziaeepour}, \& {Zschocke}}]{2016A&A...595A...1G}
{Gaia Collaboration}, {Prusti}, T., {de Bruijne}, J.~H.~J., {et~al.} 2016,
  \aap, 595, A1, \dodoi{10.1051/0004-6361/201629272}

\bibitem[{{Gaia Collaboration} {et~al.}(2018){Gaia Collaboration}, {Brown},
  {Vallenari}, {Prusti}, {de Bruijne}, {Babusiaux}, {Bailer-Jones}, {Biermann},
  {Evans}, {Eyer}, {Jansen}, {Jordi}, {Klioner}, {Lammers}, {Lindegren},
  {Luri}, {Mignard}, {Panem}, {Pourbaix}, {Randich}, {Sartoretti}, {Siddiqui},
  {Soubiran}, {van Leeuwen}, {Walton}, {Arenou}, {Bastian}, {Cropper},
  {Drimmel}, {Katz}, {Lattanzi}, {Bakker}, {Cacciari}, {Casta{\~n}eda},
  {Chaoul}, {Cheek}, {De Angeli}, {Fabricius}, {Guerra}, {Holl}, {Masana},
  {Messineo}, {Mowlavi}, {Nienartowicz}, {Panuzzo}, {Portell}, {Riello},
  {Seabroke}, {Tanga}, {Th{\'e}venin}, {Gracia-Abril}, {Comoretto},
  {Garcia-Reinaldos}, {Teyssier}, {Altmann}, {Andrae}, {Audard},
  {Bellas-Velidis}, {Benson}, {Berthier}, {Blomme}, {Burgess}, {Busso},
  {Carry}, {Cellino}, {Clementini}, {Clotet}, {Creevey}, {Davidson}, {De
  Ridder}, {Delchambre}, {Dell'Oro}, {Ducourant},
  {Fern{\'a}ndez-Hern{\'a}ndez}, {Fouesneau}, {Fr{\'e}mat}, {Galluccio},
  {Garc{\'\i}a-Torres}, {Gonz{\'a}lez-N{\'u}{\~n}ez}, {Gonz{\'a}lez-Vidal},
  {Gosset}, {Guy}, {Halbwachs}, {Hambly}, {Harrison}, {Hern{\'a}ndez},
  {Hestroffer}, {Hodgkin}, {Hutton}, {Jasniewicz}, {Jean-Antoine-Piccolo},
  {Jordan}, {Korn}, {Krone-Martins}, {Lanzafame}, {Lebzelter}, {L{\"o}ffler},
  {Manteiga}, {Marrese}, {Mart{\'\i}n-Fleitas}, {Moitinho}, {Mora}, {Muinonen},
  {Osinde}, {Pancino}, {Pauwels}, {Petit}, {Recio-Blanco}, {Richards},
  {Rimoldini}, {Robin}, {Sarro}, {Siopis}, {Smith}, {Sozzetti}, {S{\"u}veges},
  {Torra}, {van Reeven}, {Abbas}, {Abreu Aramburu}, {Accart}, {Aerts},
  {Altavilla}, {{\'A}lvarez}, {Alvarez}, {Alves}, {Anderson}, {Andrei},
  {Anglada Varela}, {Antiche}, {Antoja}, {Arcay}, {Astraatmadja}, {Bach},
  {Baker}, {Balaguer-N{\'u}{\~n}ez}, {Balm}, {Barache}, {Barata}, {Barbato},
  {Barblan}, {Barklem}, {Barrado}, {Barros}, {Barstow}, {Bartholom{\'e}
  Mu{\~n}oz}, {Bassilana}, {Becciani}, {Bellazzini}, {Berihuete}, {Bertone},
  {Bianchi}, {Bienaym{\'e}}, {Blanco-Cuaresma}, {Boch}, {Boeche}, {Bombrun},
  {Borrachero}, {Bossini}, {Bouquillon}, {Bourda}, {Bragaglia}, {Bramante},
  {Breddels}, {Bressan}, {Brouillet}, {Br{\"u}semeister}, {Brugaletta},
  {Bucciarelli}, {Burlacu}, {Busonero}, {Butkevich}, {Buzzi}, {Caffau},
  {Cancelliere}, {Cannizzaro}, {Cantat-Gaudin}, {Carballo}, {Carlucci},
  {Carrasco}, {Casamiquela}, {Castellani}, {Castro-Ginard}, {Charlot},
  {Chemin}, {Chiavassa}, {Cocozza}, {Costigan}, {Cowell}, {Crifo}, {Crosta},
  {Crowley}, {Cuypers}, {Dafonte}, {Damerdji}, {Dapergolas}, {David}, {David},
  {de Laverny}, {De Luise}, {De March}, {de Martino}, {de Souza}, {de Torres},
  {Debosscher}, {del Pozo}, {Delbo}, {Delgado}, {Delgado}, {Di Matteo},
  {Diakite}, {Diener}, {Distefano}, {Dolding}, {Drazinos}, {Dur{\'a}n},
  {Edvardsson}, {Enke}, {Eriksson}, {Esquej}, {Eynard Bontemps}, {Fabre},
  {Fabrizio}, {Faigler}, {Falc{\~a}o}, {Farr{\`a}s Casas}, {Federici},
  {Fedorets}, {Fernique}, {Figueras}, {Filippi}, {Findeisen}, {Fonti},
  {Fraile}, {Fraser}, {Fr{\'e}zouls}, {Gai}, {Galleti}, {Garabato},
  {Garc{\'\i}a-Sedano}, {Garofalo}, {Garralda}, {Gavel}, {Gavras}, {Gerssen},
  {Geyer}, {Giacobbe}, {Gilmore}, {Girona}, {Giuffrida}, {Glass}, {Gomes},
  {Granvik}, {Gueguen}, {Guerrier}, {Guiraud}, {Guti{\'e}rrez-S{\'a}nchez},
  {Haigron}, {Hatzidimitriou}, {Hauser}, {Haywood}, {Heiter}, {Helmi}, {Heu},
  {Hilger}, {Hobbs}, {Hofmann}, {Holland}, {Huckle}, {Hypki}, {Icardi},
  {Jan{\ss}en}, {Jevardat de Fombelle}, {Jonker}, {Juh{\'a}sz}, {Julbe},
  {Karampelas}, {Kewley}, {Klar}, {Kochoska}, {Kohley}, {Kolenberg},
  {Kontizas}, {Kontizas}, {Koposov}, {Kordopatis}, {Kostrzewa-Rutkowska},
  {Koubsky}, {Lambert}, {Lanza}, {Lasne}, {Lavigne}, {Le Fustec}, {Le
  Poncin-Lafitte}, {Lebreton}, {Leccia}, {Leclerc}, {Lecoeur-Taibi},
  {Lenhardt}, {Leroux}, {Liao}, {Licata}, {Lindstr{\o}m}, {Lister}, {Livanou},
  {Lobel}, {L{\'o}pez}, {Managau}, {Mann}, {Mantelet}, {Marchal}, {Marchant},
  {Marconi}, {Marinoni}, {Marschalk{\'o}}, {Marshall}, {Martino}, {Marton},
  {Mary}, {Massari}, {Matijevi{\v{c}}}, {Mazeh}, {McMillan}, {Messina},
  {Michalik}, {Millar}, {Molina}, {Molinaro}, {Moln{\'a}r}, {Montegriffo},
  {Mor}, {Morbidelli}, {Morel}, {Morris}, {Mulone}, {Muraveva}, {Musella},
  {Nelemans}, {Nicastro}, {Noval}, {O'Mullane}, {Ord{\'e}novic},
  {Ord{\'o}{\~n}ez-Blanco}, {Osborne}, {Pagani}, {Pagano}, {Pailler},
  {Palacin}, {Palaversa}, {Panahi}, {Pawlak}, {Piersimoni}, {Pineau}, {Plachy},
  {Plum}, {Poggio}, {Poujoulet}, {Pr{\v{s}}a}, {Pulone}, {Racero}, {Ragaini},
  {Rambaux}, {Ramos-Lerate}, {Regibo}, {Reyl{\'e}}, {Riclet}, {Ripepi}, {Riva},
  {Rivard}, {Rixon}, {Roegiers}, {Roelens}, {Romero-G{\'o}mez}, {Rowell},
  {Royer}, {Ruiz-Dern}, {Sadowski}, {Sagrist{\`a} Sell{\'e}s}, {Sahlmann},
  {Salgado}, {Salguero}, {Sanna}, {Santana-Ros}, {Sarasso}, {Savietto},
  {Schultheis}, {Sciacca}, {Segol}, {Segovia}, {S{\'e}gransan}, {Shih},
  {Siltala}, {Silva}, {Smart}, {Smith}, {Solano}, {Solitro}, {Sordo}, {Soria
  Nieto}, {Souchay}, {Spagna}, {Spoto}, {Stampa}, {Steele},
  {Steidelm{\"u}ller}, {Stephenson}, {Stoev}, {Suess}, {Surdej}, {Szabados},
  {Szegedi-Elek}, {Tapiador}, {Taris}, {Tauran}, {Taylor}, {Teixeira},
  {Terrett}, {Teyssandier}, {Thuillot}, {Titarenko}, {Torra Clotet}, {Turon},
  {Ulla}, {Utrilla}, {Uzzi}, {Vaillant}, {Valentini}, {Valette}, {van Elteren},
  {Van Hemelryck}, {van Leeuwen}, {Vaschetto}, {Vecchiato}, {Veljanoski},
  {Viala}, {Vicente}, {Vogt}, {von Essen}, {Voss}, {Votruba}, {Voutsinas},
  {Walmsley}, {Weiler}, {Wertz}, {Wevers}, {Wyrzykowski}, {Yoldas},
  {{\v{Z}}erjal}, {Ziaeepour}, {Zorec}, {Zschocke}, {Zucker}, {Zurbach}, \&
  {Zwitter}}]{2018A&A...616A...1G}
{Gaia Collaboration}, {Brown}, A.~G.~A., {Vallenari}, A., {et~al.} 2018, \aap,
  616, A1, \dodoi{10.1051/0004-6361/201833051}

\bibitem[{{Gaia Collaboration} {et~al.}(2022){Gaia Collaboration}, {Vallenari},
  {Brown}, {Prusti}, {de Bruijne}, {Arenou}, {Babusiaux}, {Biermann},
  {Creevey}, {Ducourant}, {Evans}, {Eyer}, {Guerra}, {Hutton}, {Jordi},
  {Klioner}, {Lammers}, {Lindegren}, {Luri}, {Mignard}, {Panem}, {Pourbaix},
  {Randich}, {Sartoretti}, {Soubiran}, {Tanga}, {Walton}, {Bailer-Jones},
  {Bastian}, {Drimmel}, {Jansen}, {Katz}, {Lattanzi}, {van Leeuwen}, {Bakker},
  {Cacciari}, {Casta{\~n}eda}, {De Angeli}, {Fabricius}, {Fouesneau},
  {Fr{\'e}mat}, {Galluccio}, {Guerrier}, {Heiter}, {Masana}, {Messineo},
  {Mowlavi}, {Nicolas}, {Nienartowicz}, {Pailler}, {Panuzzo}, {Riclet}, {Roux},
  {Seabroke}, {Sordo{\o}rcit}, {Th{\'e}venin}, {Gracia-Abril}, {Portell},
  {Teyssier}, {Altmann}, {Andrae}, {Audard}, {Bellas-Velidis}, {Benson},
  {Berthier}, {Blomme}, {Burgess}, {Busonero}, {Busso}, {C{\'a}novas}, {Carry},
  {Cellino}, {Cheek}, {Clementini}, {Damerdji}, {Davidson}, {de Teodoro},
  {Nu{\~n}ez Campos}, {Delchambre}, {Dell'Oro}, {Esquej},
  {Fern{\'a}ndez-Hern{\'a}ndez}, {Fraile}, {Garabato}, {Garc{\'\i}a-Lario},
  {Gosset}, {Haigron}, {Halbwachs}, {Hambly}, {Harrison}, {Hern{\'a}ndez},
  {Hestroffer}, {Hodgkin}, {Holl}, {Jan{\ss}en}, {Jevardat de Fombelle},
  {Jordan}, {Krone-Martins}, {Lanzafame}, {L{\"o}ffler}, {Marchal}, {Marrese},
  {Moitinho}, {Muinonen}, {Osborne}, {Pancino}, {Pauwels}, {Recio-Blanco},
  {Reyl{\'e}}, {Riello}, {Rimoldini}, {Roegiers}, {Rybizki}, {Sarro}, {Siopis},
  {Smith}, {Sozzetti}, {Utrilla}, {van Leeuwen}, {Abbas}, {{\'A}brah{\'a}m},
  {Abreu Aramburu}, {Aerts}, {Aguado}, {Ajaj}, {Aldea-Montero}, {Altavilla},
  {{\'A}lvarez}, {Alves}, {Anders}, {Anderson}, {Anglada Varela}, {Antoja},
  {Baines}, {Baker}, {Balaguer-N{\'u}{\~n}ez}, {Balbinot}, {Balog}, {Barache},
  {Barbato}, {Barros}, {Barstow}, {Bartolom{\'e}}, {Bassilana}, {Bauchet},
  {Becciani}, {Bellazzini}, {Berihuete}, {Bernet}, {Bertone}, {Bianchi},
  {Binnenfeld}, {Blanco-Cuaresma}, {Blazere}, {Boch}, {Bombrun}, {Bossini},
  {Bouquillon}, {Bragaglia}, {Bramante}, {Breedt}, {Bressan}, {Brouillet},
  {Brugaletta}, {Bucciarelli}, {Burlacu}, {Butkevich}, {Buzzi}, {Caffau},
  {Cancelliere}, {Cantat-Gaudin}, {Carballo}, {Carlucci}, {Carnerero},
  {Carrasco}, {Casamiquela}, {Castellani}, {Castro-Ginard}, {Chaoul},
  {Charlot}, {Chemin}, {Chiaramida}, {Chiavassa}, {Chornay}, {Comoretto},
  {Contursi}, {Cooper}, {Cornez}, {Cowell}, {Crifo}, {Cropper}, {Crosta},
  {Crowley}, {Dafonte}, {Dapergolas}, {David}, {David}, {de Laverny}, {De
  Luise}, {De March}, {De Ridder}, {de Souza}, {de Torres}, {del Peloso}, {del
  Pozo}, {Delbo}, {Delgado}, {Delisle}, {Demouchy}, {Dharmawardena}, {Di
  Matteo}, {Diakite}, {Diener}, {Distefano}, {Dolding}, {Edvardsson}, {Enke},
  {Fabre}, {Fabrizio}, {Faigler}, {Fedorets}, {Fernique}, {Fienga}, {Figueras},
  {Fournier}, {Fouron}, {Fragkoudi}, {Gai}, {Garcia-Gutierrez},
  {Garcia-Reinaldos}, {Garc{\'\i}a-Torres}, {Garofalo}, {Gavel}, {Gavras},
  {Gerlach}, {Geyer}, {Giacobbe}, {Gilmore}, {Girona}, {Giuffrida}, {Gomel},
  {Gomez}, {Gonz{\'a}lez-N{\'u}{\~n}ez}, {Gonz{\'a}lez-Santamar{\'\i}a},
  {Gonz{\'a}lez-Vidal}, {Granvik}, {Guillout}, {Guiraud},
  {Guti{\'e}rrez-S{\'a}nchez}, {Guy}, {Hatzidimitriou}, {Hauser}, {Haywood},
  {Helmer}, {Helmi}, {Sarmiento}, {Hidalgo}, {Hilger}, {H{\l}adczuk}, {Hobbs},
  {Holland}, {Huckle}, {Jardine}, {Jasniewicz}, {Jean-Antoine Piccolo},
  {Jim{\'e}nez-Arranz}, {Jorissen}, {Juaristi Campillo}, {Julbe}, {Karbevska},
  {Kervella}, {Khanna}, {Kontizas}, {Kordopatis}, {Korn}, {K{\'o}sp{\'a}l},
  {Kostrzewa-Rutkowska}, {Kruszy{\'n}ska}, {Kun}, {Laizeau}, {Lambert},
  {Lanza}, {Lasne}, {Le Campion}, {Lebreton}, {Lebzelter}, {Leccia}, {Leclerc},
  {Lecoeur-Taibi}, {Liao}, {Licata}, {Lindstr{\o}m}, {Lister}, {Livanou},
  {Lobel}, {Lorca}, {Loup}, {Madrero Pardo}, {Magdaleno Romeo}, {Managau},
  {Mann}, {Manteiga}, {Marchant}, {Marconi}, {Marcos}, {Marcos Santos},
  {Mar{\'\i}n Pina}, {Marinoni}, {Marocco}, {Marshall}, {Polo},
  {Mart{\'\i}n-Fleitas}, {Marton}, {Mary}, {Masip}, {Massari},
  {Mastrobuono-Battisti}, {Mazeh}, {McMillan}, {Messina}, {Michalik}, {Millar},
  {Mints}, {Molina}, {Molinaro}, {Moln{\'a}r}, {Monari}, {Mongui{\'o}},
  {Montegriffo}, {Montero}, {Mor}, {Mora}, {Morbidelli}, {Morel}, {Morris},
  {Muraveva}, {Murphy}, {Musella}, {Nagy}, {Noval}, {Oca{\~n}a}, {Ogden},
  {Ordenovic}, {Osinde}, {Pagani}, {Pagano}, {Palaversa}, {Palicio},
  {Pallas-Quintela}, {Panahi}, {Payne-Wardenaar}, {Pe{\~n}alosa Esteller},
  {Penttil{\"a}}, {Pichon}, {Piersimoni}, {Pineau}, {Plachy}, {Plum}, {Poggio},
  {Pr{\v{s}}a}, {Pulone}, {Racero}, {Ragaini}, {Rainer}, {Raiteri}, {Rambaux},
  {Ramos}, {Ramos-Lerate}, {Re Fiorentin}, {Regibo}, {Richards}, {Rios Diaz},
  {Ripepi}, {Riva}, {Rix}, {Rixon}, {Robichon}, {Robin}, {Robin}, {Roelens},
  {Rogues}, {Rohrbasser}, {Romero-G{\'o}mez}, {Rowell}, {Royer}, {Ruz Mieres},
  {Rybicki}, {Sadowski}, {S{\'a}ez N{\'u}{\~n}ez}, {Sagrist{\`a} Sell{\'e}s},
  {Sahlmann}, {Salguero}, {Samaras}, {Sanchez Gimenez}, {Sanna},
  {Santove{\~n}a}, {Sarasso}, {Schultheis}, {Sciacca}, {Segol}, {Segovia},
  {S{\'e}gransan}, {Semeux}, {Shahaf}, {Siddiqui}, {Siebert}, {Siltala},
  {Silvelo}, {Slezak}, {Slezak}, {Smart}, {Snaith}, {Solano}, {Solitro},
  {Souami}, {Souchay}, {Spagna}, {Spina}, {Spoto}, {Steele},
  {Steidelm{\"u}ller}, {Stephenson}, {S{\"u}veges}, {Surdej}, {Szabados},
  {Szegedi-Elek}, {Taris}, {Taylo}, {Teixeira}, {Tolomei}, {Tonello}, {Torra},
  {Torra}, {Torralba Elipe}, {Trabucchi}, {Tsounis}, {Turon}, {Ulla}, {Unger},
  {Vaillant}, {van Dillen}, {van Reeven}, {Vanel}, {Vecchiato}, {Viala},
  {Vicente}, {Voutsinas}, {Weiler}, {Wevers}, {Wyrzykowski}, {Yoldas}, {Yvard},
  {Zhao}, {Zorec}, {Zucker}, \& {Zwitter}}]{2022arXiv220800211G}
{Gaia Collaboration}, {Vallenari}, A., {Brown}, A.~G.~A., {et~al.} 2022, arXiv
  e-prints, arXiv:2208.00211.
\newblock \doarXiv{2208.00211}

\bibitem[{Green {et~al.}(2019)Green, Schlafly, Zucker, Speagle, \&
  Finkbeiner}]{Green_2019}
Green, G.~M., Schlafly, E., Zucker, C., Speagle, J.~S., \& Finkbeiner, D. 2019,
  The Astrophysical Journal, 887, 93, \dodoi{10.3847/1538-4357/ab5362}

\bibitem[{{Hellier}(1993)}]{1993MNRAS.264..132H}
{Hellier}, C. 1993, \mnras, 264, 132, \dodoi{10.1093/mnras/264.1.132}

\bibitem[{{Hoard} {et~al.}(2003){Hoard}, {Szkody}, {Froning}, {Long}, \&
  {Knigge}}]{2003AJ....126.2473H}
{Hoard}, D.~W., {Szkody}, P., {Froning}, C.~S., {Long}, K.~S., \& {Knigge}, C.
  2003, \aj, 126, 2473, \dodoi{10.1086/378605}

\bibitem[{{Hoard} {et~al.}(2000){Hoard}, {Thorstensen}, \&
  {Szkody}}]{2000ApJ...537..936H}
{Hoard}, D.~W., {Thorstensen}, J.~R., \& {Szkody}, P. 2000, \apj, 537, 936,
  \dodoi{10.1086/309074}

\bibitem[{{Hou} {et~al.}(2020){Hou}, {Luo}, {Li}, \&
  {Qin}}]{2020AJ....159...43H}
{Hou}, W., {Luo}, A.~l., {Li}, Y.-B., \& {Qin}, L. 2020, \aj, 159, 43,
  \dodoi{10.3847/1538-3881/ab5962}

\bibitem[{{Huang} {et~al.}(2020){Huang}, {Vanderburg}, {P{\'a}l}, {Sha}, {Yu},
  {Fong}, {Fausnaugh}, {Shporer}, {Guerrero}, {Vanderspek}, \&
  {Ricker}}]{2020RNAAS...4..204H}
{Huang}, C.~X., {Vanderburg}, A., {P{\'a}l}, A., {et~al.} 2020, Research Notes
  of the American Astronomical Society, 4, 204,
  \dodoi{10.3847/2515-5172/abca2e}

\bibitem[{{Huang} {et~al.}(2012){Huang}, {Li}, {Wang}, {Shang}, {Zhang}, {Hu},
  {Qiu}, \& {Jiang}}]{2012RAA....12.1585H}
{Huang}, F., {Li}, J.-Z., {Wang}, X.-F., {et~al.} 2012, Research in Astronomy
  and Astrophysics, 12, 1585, \dodoi{10.1088/1674-4527/12/11/012}

\bibitem[{{I{\l}kiewicz} {et~al.}(2021){I{\l}kiewicz}, {Scaringi}, {Court},
  {Maccarone}, {Altamirano}, {Bradshaw}, {Degenaar}, {Fratta}, {Littlefield},
  {Shahbaz}, \& {Wijnands}}]{2021MNRAS.503.4050I}
{I{\l}kiewicz}, K., {Scaringi}, S., {Court}, J. M.~C., {et~al.} 2021, \mnras,
  503, 4050, \dodoi{10.1093/mnras/stab664}

\bibitem[{{King}(1988)}]{1988QJRAS..29....1K}
{King}, A.~R. 1988, \qjras, 29, 1

\bibitem[{Kochanek {et~al.}(2017)Kochanek, Shappee, Stanek, Holoien, Thompson,
  Prieto, Dong, Shields, Will, Britt, \& et~al.}]{Kochanek_2017}
Kochanek, C.~S., Shappee, B.~J., Stanek, K.~Z., {et~al.} 2017, \pasp, 129,
  104502, \dodoi{10.1088/1538-3873/aa80d9}

\bibitem[{{Kolb} {et~al.}(1998){Kolb}, {King}, \&
  {Ritter}}]{1998MNRAS.298L..29K}
{Kolb}, U., {King}, A.~R., \& {Ritter}, H. 1998, \mnras, 298, L29,
  \dodoi{10.1046/j.1365-8711.1998.01854.x}

\bibitem[{Lima {et~al.}(2021)Lima, Rodrigues, Lopes, Szkody, Jablonski,
  Oliveira, Silva, Belloni, Palhares, Shugarov, Baptista, \&
  Almeida}]{Lima_2021}
Lima, I.~J., Rodrigues, C.~V., Lopes, C. E.~F., {et~al.} 2021, The Astronomical
  Journal, 161, 225, \dodoi{10.3847/1538-3881/abeb16}

\bibitem[{{Lomb}(1976)}]{1976Ap&SS..39..447L}
{Lomb}, N.~R. 1976, \apss, 39, 447, \dodoi{10.1007/BF00648343}

\bibitem[{{Lubow} \& {Pringle}(1993)}]{1993ApJ...409..360L}
{Lubow}, S.~H., \& {Pringle}, J.~E. 1993, \apj, 409, 360,
  \dodoi{10.1086/172669}

\bibitem[{{Margoni} \& {Stagni}(1984)}]{1984A&AS...56...87M}
{Margoni}, R., \& {Stagni}, R. 1984, \aaps, 56, 87

\bibitem[{{Osaki}(1989)}]{1989PASJ...41.1005O}
{Osaki}, Y. 1989, \pasj, 41, 1005

\bibitem[{{Patterson}(1979)}]{1979AJ.....84..804P}
{Patterson}, J. 1979, \aj, 84, 804, \dodoi{10.1086/112483}

\bibitem[{{Patterson}(1995)}]{1995PASP..107..657P}
---. 1995, \pasp, 107, 657, \dodoi{10.1086/133605}

\bibitem[{{Patterson} {et~al.}(1993){Patterson}, {Halpern}, \&
  {Shambrook}}]{1993ApJ...419..803P}
{Patterson}, J., {Halpern}, J., \& {Shambrook}, A. 1993, \apj, 419, 803,
  \dodoi{10.1086/173532}

\bibitem[{{Patterson} \& {Skillman}(1994)}]{1994PASP..106.1141P}
{Patterson}, J., \& {Skillman}, D.~R. 1994, \pasp, 106, 1141,
  \dodoi{10.1086/133491}

\bibitem[{{Patterson} {et~al.}(2002){Patterson}, {Fenton}, {Thorstensen},
  {Harvey}, {Skillman}, {Fried}, {Monard}, {O'Donoghue}, {Beshore}, {Martin},
  {Niarchos}, {Vanmunster}, {Foote}, {Bolt}, {Rea}, {Cook}, {Butterworth}, \&
  {Wood}}]{2002PASP..114.1364P}
{Patterson}, J., {Fenton}, W.~H., {Thorstensen}, J.~R., {et~al.} 2002, \pasp,
  114, 1364, \dodoi{10.1086/344587}

\bibitem[{{Patterson} {et~al.}(2013){Patterson}, {Uthas}, {Kemp}, {de Miguel},
  {Krajci}, {Foote}, {Hambsch}, {Campbell}, {Roberts}, {Cejudo}, {Dvorak},
  {Vanmunster}, {Koff}, {Skillman}, {Harvey}, {Martin}, {Rock}, {Boyd},
  {Oksanen}, {Morelle}, {Ulowetz}, {Kroes}, {Sabo}, \&
  {Jensen}}]{2013MNRAS.434.1902P}
{Patterson}, J., {Uthas}, H., {Kemp}, J., {et~al.} 2013, \mnras, 434, 1902,
  \dodoi{10.1093/mnras/stt1085}

\bibitem[{Ricker {et~al.}(2014)Ricker, Winn, Vanderspek, Latham, Bakos, Bean,
  Berta-Thompson, Brown, Buchhave, Butler, Butler, Chaplin, Charbonneau,
  Christensen-Dalsgaard, Clampin, Deming, Doty, Lee, Dressing, Dunham, Endl,
  Fressin, Ge, Henning, Holman, Howard, Ida, Jenkins, Jernigan, Johnson,
  Kaltenegger, Kawai, Kjeldsen, Laughlin, Levine, Lin, Lissauer, MacQueen,
  Marcy, McCullough, Morton, Narita, Paegert, Palle, Pepe, Pepper, Quirrenbach,
  Rinehart, Sasselov, Sato, Seager, Sozzetti, Stassun, Sullivan, Szentgyorgyi,
  Torres, Udry, \& Villasenor}]{10.1117/1.JATIS.1.1.014003}
Ricker, G.~R., Winn, J.~N., Vanderspek, R., {et~al.} 2014, Journal of
  Astronomical Telescopes, Instruments, and Systems, 1, 1 ,
  \dodoi{10.1117/1.JATIS.1.1.014003}

\bibitem[{{Scargle}(1982)}]{1982ApJ...263..835S}
{Scargle}, J.~D. 1982, \apj, 263, 835, \dodoi{10.1086/160554}

\bibitem[{{Shafter} {et~al.}(1990){Shafter}, {Robinson}, {Crampton}, {Warner},
  \& {Prestage}}]{1990ApJ...354..708S}
{Shafter}, A.~W., {Robinson}, E.~L., {Crampton}, D., {Warner}, B., \&
  {Prestage}, R.~M. 1990, \apj, 354, 708, \dodoi{10.1086/168727}

\bibitem[{{Stefanov} {et~al.}(2022){Stefanov}, {Latev}, {Boeva}, \&
  {Moyseev}}]{2022MNRAS.tmp.2239S}
{Stefanov}, S.~Y., {Latev}, G., {Boeva}, S., \& {Moyseev}, M. 2022, \mnras,
  \dodoi{10.1093/mnras/stac2317}

\bibitem[{Szkody {et~al.}(2020)Szkody, Dicenzo, Ho, Hillenbrand, van Roestel,
  Ridder, DeJesus~Lima, Graham, Bellm, Burdge, \& et~al.}]{Szkody_2020}
Szkody, P., Dicenzo, B., Ho, A. Y.~Q., {et~al.} 2020, \aj, 159, 198,
  \dodoi{10.3847/1538-3881/ab7cce}

\bibitem[{Szkody {et~al.}(2021)Szkody, Olde~Loohuis, Koplitz, van Roestel,
  Dicenzo, Ho, Hillenbrand, Bellm, Dekany, Drake, \& et~al.}]{Szkody_2021}
Szkody, P., Olde~Loohuis, C., Koplitz, B., {et~al.} 2021, \aj, 162, 94,
  \dodoi{10.3847/1538-3881/ac0efb}

\bibitem[{{Thorstensen}(2020)}]{2020AJ....160..151T}
{Thorstensen}, J.~R. 2020, \aj, 160, 151, \dodoi{10.3847/1538-3881/aba7c7}

\bibitem[{{Udalski} \& {Schwarzenberg-Czerny}(1989)}]{1989AcA....39..125U}
{Udalski}, A., \& {Schwarzenberg-Czerny}, A. 1989, \actaa, 39, 125

\bibitem[{{VanderPlas}(2018)}]{2018ApJS..236...16V}
{VanderPlas}, J.~T. 2018, \apjs, 236, 16, \dodoi{10.3847/1538-4365/aab766}

\bibitem[{{Wang} {et~al.}(1996){Wang}, {Su}, {Chu}, {Cui}, \&
  {Wang}}]{1996ApOpt..35.5155W}
{Wang}, S.-G., {Su}, D.-Q., {Chu}, Y.-Q., {Cui}, X., \& {Wang}, Y.-N. 1996,
  \ao, 35, 5155, \dodoi{10.1364/AO.35.005155}

\bibitem[{Wang {et~al.}(2008)Wang, Li, Filippenko, Krisciunas, Suntzeff, Li,
  Zhang, Deng, Foley, Ganeshalingam, Li, Lou, Qiu, Shang, Silverman, Zhang, \&
  Zhang}]{Wang_2008}
Wang, X., Li, W., Filippenko, A.~V., {et~al.} 2008, The Astrophysical Journal,
  675, 626, \dodoi{10.1086/526413}

\bibitem[{Warner(1995)}]{warner_1995}
Warner, B. 1995, Cataclysmic Variable Stars, Cambridge Astrophysics (Cambridge
  University Press), \dodoi{10.1017/CBO9780511586491}

\bibitem[{{Wood} {et~al.}(2009){Wood}, {Thomas}, \&
  {Simpson}}]{2009MNRAS.398.2110W}
{Wood}, M.~A., {Thomas}, D.~M., \& {Simpson}, J.~C. 2009, \mnras, 398, 2110,
  \dodoi{10.1111/j.1365-2966.2009.15252.x}

\bibitem[{{Woudt} \& {Warner}(2002)}]{2002MNRAS.333..411W}
{Woudt}, P.~A., \& {Warner}, B. 2002, \mnras, 333, 411,
  \dodoi{10.1046/j.1365-8711.2002.05415.x}

\bibitem[{{Wu} {et~al.}(2002){Wu}, {Li}, {Ding}, {Zhang}, \&
  {Li}}]{2002ApJ...569..418W}
{Wu}, X., {Li}, Z., {Ding}, Y., {Zhang}, Z., \& {Li}, Z. 2002, \apj, 569, 418,
  \dodoi{10.1086/339278}

\end{thebibliography}
\bibliographystyle{aasjournal}

\end{CJK*}
\end{document}